\documentclass{aa}

\usepackage{graphicx}
\usepackage{txfonts}
\usepackage{multirow}
\begin{document}

   \title{Vertical CO surfaces as a probe for\\ protoplanetary disk mass and carbon depletion}
    \authorrunning{T. Paneque-Carre\~no et al.}
    \author{T. Paneque-Carre\~no \inst{1,2}, A. Miotello \inst{1}, E. F. van Dishoeck \inst{2, 3}, G. Rosotti \inst{4}, B. Tabone \inst{5}
            }

    \institute{European Southern Observatory, Karl-Schwarzschild-Str 2, 85748 Garching, Germany
    \and Leiden Observatory, Leiden University, P.O. Box 9513, NL-2300 RA Leiden, the Netherlands
    \and Max-Planck-Institut für extraterrestrische Physik, Gießenbachstr. 1 , 85748 Garching bei München, Germany
    \and Dipartimento di Fisica, Università degli Studi di Milano, Via Celoria, 16, Milano, I-20133, Italy
    \and Universit\'e Paris-Saclay, CNRS, Institut d’Astrophysique Spatiale, 91405 Orsay, France
    \\
              \email{tpaneque@umich.edu}}
   \date{}


  \abstract
   {As the sample of mid-inclination disks with measured CO emission surfaces grows, a fundamental unanswered question is how these vertical profiles connect to their host properties.}
   {This project aims to relate the vertical extent of protoplanetary disks as traced by $^{12}$CO $2-1$ to key stellar and physical parameters. In order to produce a result that is applicable towards an observational analysis, we benchmark our results with ALMA observations of CO emission from nineteen disks.}
   {We produce a grid of disk models using the physical-chemical code DALI, for a template T\,Tauri and Herbig star. Our models use an iterative solver to calculate the hydrostatic equilibrium equations and determine a physically-motivated density structure. Key stellar and disk parameters such as stellar luminosity and temperature, total disk mass, carbon abundance and critical radius are varied to determine their effect on the CO emitting surface. Each vertical profile is fitted by an exponentially tapered power-law and characterized by the $z/r$ value that represents the structure inwards of 80\% of the tapering radius.}
   {The CO emission surface location is primarily determined by the disk mass ($M_d$) and the level of volatile carbon depletion. T\,Tauri and Herbig systems show different vertical profiles, with disks around T\,Tauri stars being more vertically extended. We derive a $z/r$-$M_d$ relationship, which for each stellar type has a degeneracy with the volatile carbon abundance. In order to reconcile total disk mass estimates from the characteristic $z/r$ and the values obtained based on dust continuum analysis, a volatile carbon depletion of 10-100 (with respect to the ISM) is needed for the majority of our sources. Our carbon depletion values are in agreement with previous literature estimates, highlighting the potential of this method to rapidly calculate key disk parameters.}
   {}


   \maketitle
%

\section{Introduction}

Planets form and grow within protoplanetary disks made of dust and gas which orbit around young stellar objects. Through high spatial and spectral resolutions of the Atacama Large Millimeter/submillimeter Array (ALMA) it has been possible to trace the extent and structure of these disks in dust and gas across all three dimensions; azimuthal, radial and vertical \citep[e.g.][]{DSHARP_Huang_radial, Andrews_2020_review, MAPS_Law_radial, MAPS_Law_Surf, Paneque_2023_vert, Temmink_2023_HD142527, booth_2023_hd100546}. To date, most studies have focused on the information offered by the radial and azimuthal structure \citep[for a review see][]{Bae_2023_ppvii}. Radial extent compared to the stellar age is used as an indicator on the governing evolution processes \citep{Rosotti_2019, Tazzari_2021_dustsizes, Long_2022_disksize, Trapman_2023_disksize} and radial substructures characterize a number of chemical and physical scenarios \citep{Andrews_2020_review, MAPS_Oberg, Oberg_2023_chemreview}. Azimuthal enhancements can signal towards the presence of planetary companions \citep{Ragusa_2017_horseshoes, Perez_2020_hd100546, Long_2022_lagrangian, booth_2023_hd100546, Rampinelli_2024_PDS70} and some azimuthal asymmetries can be linked to inner warps or other disk processes such as ongoing infall \citep{Perez_2018_DSHARP, Young_2021_warp, Paneque-Carreno_2022_Elias_CN}. Vertical structure, however, has been less accessible from an observational point of view and therefore less studied in comparison to the other two dimensions.

Directly probing the vertical structure and molecular layering of disks through observations of edge-on disks has offered valuable insight on the thermal and density conditions \citep{Dutrey_2017_flyingsaucer, RuizRodriguez_2021}, as well as dust-settling processes \citep{Villenave_2020, Villenave_2023}. In the recent years, novel methodologies \citep{Dartois_2003_DMTau, Pinte_2018_method, Kurtovic_2024_MHO6} have proven that the vertical structure can also be traced from mid-inclination ($\sim$40-60$^{\circ}$) disks, opening an exciting avenue to explore this dimension in larger samples \citep[e.g.][]{Rich_2021, MAPS_Law_Surf, Law_2022_12CO, Law_2023_CO_isotop_surf} and multiple molecular tracers \citep[e.g.][]{Paneque-Carreno_2022_Elias_CN,Paneque_2023_vert,paneque-carreno_2024_IMLup, Law_2023_CI, Urbina_2024_CI, Hernandez-Vera_2024_H2CO_HD16}. From a theoretical point of view, the vertical distribution of the gaseous disk material will follow hydrostatic equilibrium between the stellar gravitational potential and thermal pressure support \citep{Armitage_2015}. This implies that a number of properties such as stellar mass and radiation field, but also disk temperature structure and density distribution are expected to affect the vertical extent \citep{Dalessio_1998, Aikawa_2002}. While the bulk of the disk material is composed of molecular hydrogen (H$_2$), we are not able to trace the disk structure through its emission, as it is extremely hard to detect throughout the cold outer ($r>$50\,au) disk region. Due to this difficulty, less abundant, but bright tracers, such as carbon monoxide (CO) are preferred to characterize the disk structure and conditions \citep[for a review see][]{Miotello_ppvii_2023}.

This work focuses on the protoplanetary disk properties that affect the vertical structure as traced by bright $^{12}$CO $2-1$ emission. We select this specific transition as it is easily observed by ALMA in protoplanetary disks and the most common tracer of the disk vertical extent \citep[e.g.][]{Law_2022_12CO, Law_2023_CO_isotop_surf}. CO is the main carrier of gas-phase carbon in the interstellar medium and its chemistry is relatively simple and reliably implemented in various physical-chemical models \citep[e.g.][]{Aikawa_2002, Gorti_2008_CO, Bruderer_DALI_2013,  Miotello_2014, woitke_2016_prodimo, Yu_2016_CO, Molyarova_2017_CO}. For this reason CO is a preferred molecule to study the disk extent \citep[radially and vertically,][]{Ansdell_2018, Sanchis_2021, MAPS_Law_radial, Law_2022_12CO, Rich_2021}, kinematical processes \citep{Pinte_2023_ppvii, Izquierdo_2023_discminer2, MAPS_Teague}, temperature structure \citep{MAPS_Law_Surf, Paneque_2023_vert, Stapper_2023_Herbig_surf} and disk mass \citep[which usually requires analysis of rarer CO isotopologues,][]{Miotello_2014, miotello_2016, BoothA_2019_13C17O, stapper_2024_gasmass}. Vertically, CO gas emission originates from the warm molecular layer defined by the boundary of UV photo-dissociation in the upper regions and low temperatures that cause CO freeze-out ($\sim$20\,K) closer to the midplane \citep{Aikawa_2002}. At typical temperatures (40–60 K) and CO column densities ($\geq$ 5 $\times$ 10$^{16}$ cm$^{-2}$) of the molecular layer in protoplanetary disks, the CO millimeter emission will be optically thick \citep[$\tau \geq$ 1, ][]{van_der_Tak_2007_RADEX}. Indeed, previous works have shown that the emitting surface traced by CO, as extracted from line emission channel maps, follows the CO $\tau \sim$1 region of the disk structure \citep{Paneque_2023_vert}. Efforts in linking the observationally derived CO surfaces to system or emission properties such as stellar mass, brightness temperature or disk radius have only provided tentative trends \citep{Law_2022_12CO, Law_2023_CO_isotop_surf} and it is not yet well understood how the wide range of CO vertical profiles \citep[with z/r values of $\sim$0.1-0.5,][]{Law_2023_CO_isotop_surf, Paneque_2023_vert} relates to the system properties.

A crucial assumption that most studies related to CO emission make is that its abundance with respect to H$_2$, as set by the volatile carbon abundance, follows the canonical interstellar medium (ISM) value of $\sim$10$^{-4}$ \citep[e.g.][]{Andrews_2011, Ansdell_2016_Lupus, Flaherty_2015_hd16_weakturb, stapper_2024_gasmass}. This is in tension with observational results that show that between 1-2 orders of magnitude of volatile carbon depletion must be present to account for disk masses derived from HD observations \citep{Favre_2013_lowCO, McClure_2016_HD, Schwarz_2016_CO_TWHya, Calahan_2021_HD_TwHya} and derivations of CO depletion, with respect to ISM abundance, from optically thin CO isotopologue emission profiles \citep{Zhang_2019_CO_abundance, MAPS_Zhang}. Lower-than-ISM CO abundances may be caused by chemical reprocessing of CO into other molecules \citep{Bergin_2014_CO, Bosman_2018_CO, Diop_2024_CO_ML} or physical processing through which carbon is removed from the disk atmosphere by icy dust grains that grow and settle towards the disk midplane \citep{Krijt_2016, Krijt_2018, Krijt_2020}. These processes are in addition to photodissociation and freeze-out, which will also lower the CO abundance throughout the disk and are included in standard thermochemical models \citep[e.g.,][]{Bruderer_DALI_2013,  woitke_2016_prodimo}. Overall, it is expected that the CO abundance may change as the disk evolves and therefore it is not inherited from the natal cloud environment \citep{Visser_2009_CO, Bergner_2020_CO, MAPS_Zhang}. To the best of our knowledge, there are no previous studies that model or observationally analyze the effect of volatile carbon depletion on the vertical emitting surface of disks.

The remainder of this paper is organized as follows. Section 2 details the setup of thermochemical models used in our analysis, indicating the relevance of accounting for hydrostatic equilibrium for a more realistic approach. Section 3 shows our main results, an overview of the sampled parameter space and the effect of various disk and stellar properties on the location of the $\tau =$ 1 CO emission layer. We present a $z/r-$Disk mass relation for both T\,Tauri and Herbig disk systems and apply it to a sample of archival protoplanetary disk observations. Our results showcase the necessity of considering carbon depletion in order to obtain disk masses that are in agreement with literature values. Section 4 has a detailed discussion of our results, comparing them to previous studies and literature values of carbon depletion. Finally, Section 5 indicates the main conclusions of this work.

\section{Modelling}

The results shown in this paper are based on an extensive grid of models run with the thermochemical code DALI \citep{Bruderer_DALI_2013}. For details on the code, how the chemical network is setup and the calculation of the temperature structure, dust populations and flux values we refer to past works using the same code \citep[e.g.][]{Bruderer_2012, Bruderer_DALI_2013, miotello_2016, leemker_2022_LkCa15, stapper_2024_gasmass}. Our models are run using the chemical network from \citet{Miotello_2014}, that accounts for CO isotope-selective effects, considering selective photodissociation, fractionation reactions, self-shielding, and freeze-out \citep[for more details see][]{Miotello_2014, miotello_2016}.

\subsection{Initial DALI setup}

\begin{table}
\caption{Initial parameters in fiducial DALI model}
\label{table_DALI_param}
\setlength{\tabcolsep}{5pt}
\def\arraystretch{1.3}
\centering
\begin{tabular}{l l}
\hline\hline
Parameter & Value \\
\hline
    \multicolumn{2}{l}{\textit{Chemistry}}\\
    Chemical Age & 1\,Myr \\
   Volatile {[C]}/{[H]}  & 1.35 $\times$ 10$^{-4}$, 10$^{-5}$, 10$^{-6}$  \\
   Volatile {[O]}/{[H]}  & 2.88 $\times$ 10$^{-4}$, 10$^{-5}$, 10$^{-6}$  \\
   {[PAH]} &  10$^{-4}$ ISM abundance\\
\hline
    \multicolumn{2}{l}{\textit{Physical Structure}}\\
   gas-to-dust ratio & 100 \\
   $f_\mathrm{{large}}$ (small/large grains) & 0.95  \\
   $\chi$ (settling of large grains) & 0.2 \\
   $\varphi$ (flaring index) & 0.1\\
   $h_{100}$ (scale height aspect ratio) & 0.1\\
   $R_c$ (characteristic radii) & 100\,au \\
   $M_d$ (total disk mass) & 10$^{-1}$, 10$^{-2}$, 10$^{-3}$, 10$^{-4}$\,M$_{\odot}$ \\
\hline
    \multicolumn{2}{l}{\textit{Stellar Parameters}}\\
    $L_x$ & 1.0 $\times$ 10$^{30}$ erg\,s$^{-1}$\\
    $T_x$ & 7.0 $\times$ 10$^{7}$ K \\
\hline
\multicolumn{2}{l}{}\\
\hline
    \multicolumn{2}{l}{\textit{T\,Tauri model}}\\
    $M_*$ & 1.0 M$_{\odot}$\\
    $\dot{M}_{\mathrm{acc}}$ & 10$^{-8}$\,M$_{\odot}$\,yr$^{-1}$\\
    $L_{\mathrm{bol}}$ & 1.0 L$_{\odot}$\\
    $T_{\mathrm{eff}}$ & 4000\,K\\
\hline
\multicolumn{2}{l}{}\\
\hline
    \multicolumn{2}{l}{\textit{Herbig model}}\\
    $M_*$ & 2.0 M$_{\odot}$\\
    $L_{\mathrm{bol}}$ & 17.0 L$_{\odot}$\\
    $T_{\mathrm{eff}}$ & 9000\,K\\

\hline
\end{tabular}
\end{table}

The initial setup of the radial structure for the gas and dust in DALI follows the self-similar solution to a viscously evolving disk \citep{Lynden_Bell_Pringle_1974, Andrews_2011}.

\begin{equation}
    \Sigma(R) = \Sigma_c \left( \frac{R}{R_c} \right) ^{-\gamma} \mathrm{exp} \left[ - \left(\frac{R}{R_c} \right) ^{2-\gamma} \right]
\end{equation}

where $R_c$ is the characteristic radius at which the surface density is $\Sigma_c/e$ and $\gamma$ the surface density exponent. The initial vertical structure of the disk follows a Gaussian distribution, where the gas density ($\rho$) corresponds to,

\begin{equation}
	\rho (z) = \rho_c (r) \mathrm{exp} \left[-\frac{z^{2}}{(2 H^2)} \right]
\end{equation}

and $H$ is the disk scale height and $\rho_c$ a characteristic gas density. To set the scale heigh aspect ratio ($H/r$), we have modified the usual DALI prescription \citep[where the aspect ratio is typically noted as $h$, see for example][]{leemker_2022_LkCa15} to match the parametrization of \citet{MAPS_Zhang}. In our model the scale height aspect ratio as a function of radius follows,

\begin{equation}
    H/r = h_\mathrm{100} \left( \frac{R}{ 100 \mathrm{au} } \right) ^{\varphi}
\end{equation}

where $h_\mathrm{100}$ is the scale height aspect ratio at 100\,au and $\varphi$ the flaring angle. The main results in this study are based on the fiducial models for a T\,Tauri and Herbig system following the values presented in Table \ref{table_DALI_param}. The parameters of the T\,Tauri system are based on the models of \citet{miotello_2016}, while the fiducial Herbig model has the stellar parameters (mass, luminosity and temperature) based on HD\,163296 \citep{MAPS_Oberg, Paneque_2023_vert}. For the T\,Tauri model, excess UV radiation due to a mass accretion rate of 10$^{-8}$\,M$_{\odot}$\,yr$^{-1}$ was taken into account. It was assumed that the gravitational potential energy of the accreted mass is released with 100\% efficiency as blackbody emission with $T = 10^{4}$\,K \citep{Miotello_2014, miotello_2016}. For our parameters, these assumptions result in $L_{\mathrm{FUV}}$/$L_{\mathrm{bol}} = 1.46 \times 10^{-1}$. No accretion is considered in the Herbig system, based on observational evidence \citep{Hartmann_2016} and previous models \citep{stapper_2024_gasmass}. Beyond the fiducial models, we study a wider parameter space to determine the effect of key disk properties on the location of the vertical structure as traced by $^{12}$CO. Table \ref{table_DALI_varyparam} indicates the sampled values which includes a range of disk masses, stellar luminosities and temperatures, to cover a broader observational sample. The fraction of small/large dust grains is also an important number that may modify the CO emitting layer as it affects the degree of penetrating energetic radiation responsible for enhancing or inhibiting CO photo-dissociation.

\begin{table}
\caption{Variation of DALI model parameters. Bold numbers indicate the fiducial model values.}
\label{table_DALI_varyparam}
\setlength{\tabcolsep}{5pt}
\def\arraystretch{1.3}
\centering
\begin{tabular}{l l}
\hline\hline
Parameter & Sampled values \\
\hline
    \multicolumn{2}{l}{\textit{T\,Tauri model}}\\
    $\varphi$ (flaring index) & \textbf{0.1}, 0.2, 0.5\\
    $f_\mathrm{{large}}$ (small/large grains) & 0.9, \textbf{0.95}, 0.99\\
    $T_{\mathrm{eff}}$ & 3500, \textbf{4000}, 4500\,K\\
    $L_{\mathrm{bol}}$ & 0.2, \textbf{1.0}, 3\,L$_{\odot}$\\
    $\dot{M}_{\mathrm{acc}}$ & 10$^{-9}$, \textbf{10$^{-8}$}, 10$^{-7}$\,M$_{\odot}$\,yr$^{-1}$\\
    $R_c$ (characteristic radii) & 15, 50, \textbf{100}\,au \\
\hline
\multicolumn{2}{l}{}\\
\hline
    \multicolumn{2}{l}{\textit{Herbig model}} \\
    {[PAH]} &   \textbf{10$^{-4}$}, 10$^{-2}$ , 10$^{-1}$  ISM abundance\\
    $T_{\mathrm{eff}}$ & 7500, \textbf{9000}, 10000\,K \\
    $L_{\mathrm{bol}}$ & 10, \textbf{17}, 30\,L$_{\odot}$ \\
    $R_c$ (characteristic radii) & 15, 50, \textbf{100}\,au \\

\hline
\end{tabular}
\end{table}

\subsection{Hydrostatic equilibrium}

\begin{figure*}[h!]
   \centering
   \includegraphics[width=\hsize]{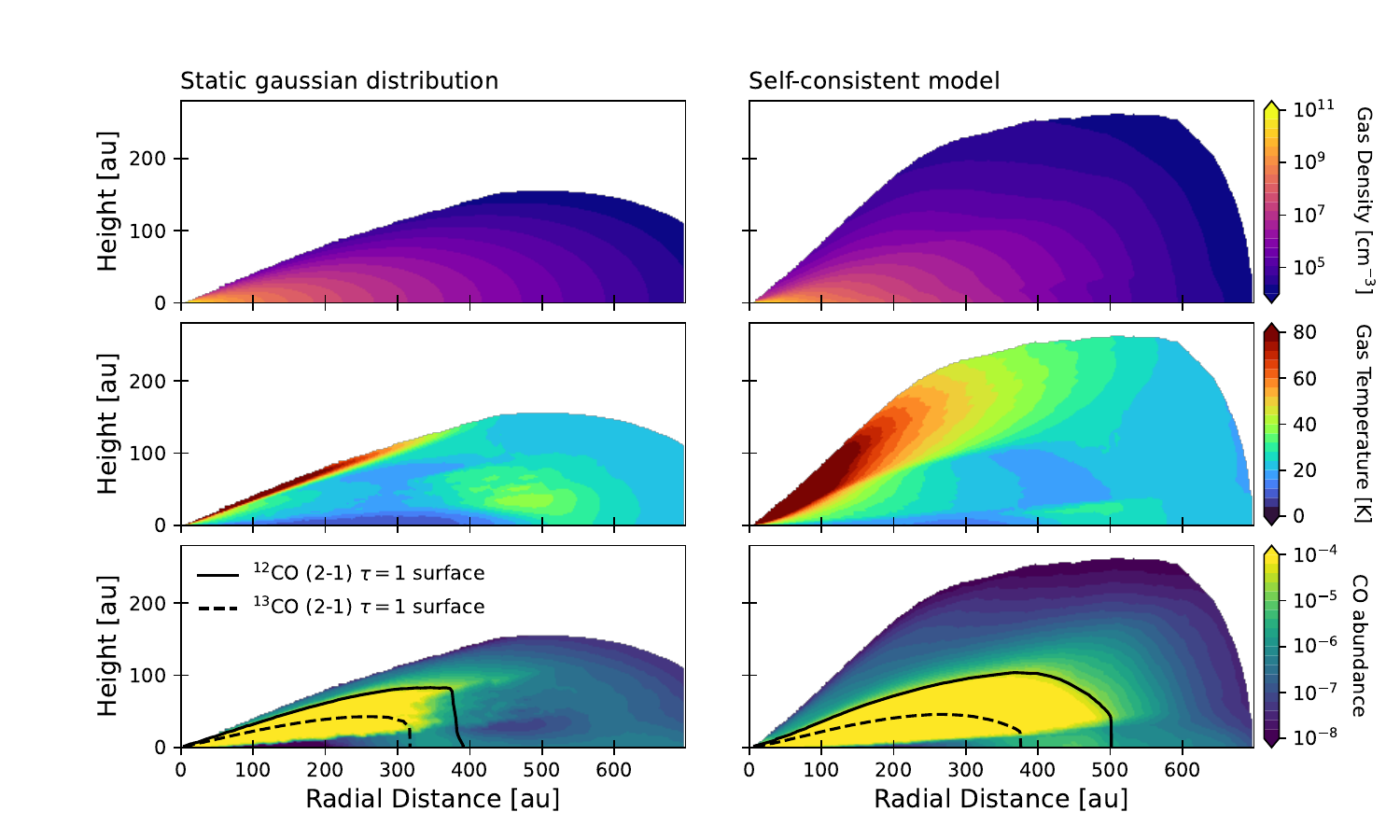}
      \caption{Comparison between the outputs of a vertically static gaussian distribution (left column) and a model after solving self-consistently for the vertical material distribution (right column). Each row shows the 2D distribution of key parameters, displayed using the same colorbar. From top to bottom the volumetric gas density, gas temperature and CO abundance (relative to H$_2$) are shown. In the bottom row the solid line traces the $^{12}$CO $\tau = $ 1 surface and the dashed line the $^{13}$CO $\tau = $ 1 surface. These outputs are from our fiducial T\,Tauri star model with a 10$^{-2}$\,M$_{\odot}$ disk.
              }
         \label{ttauri_model}
\end{figure*}

The initial setup presented in the previous section is a useful approximation, but is not realistic, as it is known that the disk vertical structure does not follow a gaussian distribution, rather it is set by hydrostatic equilibrium between the stellar gravity and pressure support of the disk (given by the thermal structure). To produce more realistic models, DALI has a hydrostatic equilibrium solver implemented, which self-consistently iterates over consecutive models, starting from the initial parametrization, to compute the disk structure. We use this approach for all of our analysis. To this end, the hydrostatic equilibrium equation that must be solved in the vertical direction, using cylindrical coordinates, is:

\begin{equation}
   \frac{1}{\rho}\frac{dP}{dz} = -\frac{d\phi}{dz}
\end{equation}

where $\rho$ is the gas density, $P$ the gas pressure and $\phi$ the gravitational potential. Without considering self-gravity, $\phi$ corresponds to the stellar gravitational potential and follows,

\begin{equation}
   \phi = -\frac{GM_*}{(r^2 + z^2)^{1/2}}
\end{equation}

Self-gravity may become relevant for the most massive disks in our parameter space, producing lower emitting layers than our prescription. However, its effect is prevalent in the outer portions of the disk \citep[r $\gtrsim R_c$,][]{Lodato_2023_mass_imlup} which are not relevant for the $z/r$ calculations of this study (method detailed in section 3).

Pressure and density can be connected such that $P/\rho = c_{s}^{2} = kT/(\mu m_p)$. Disks are not isothermal in their vertical direction, therefore, the sound speed ($c_s$) will vary as function of height, depending on the density and thermal conditions. The differential equation for $P$ is

\begin{equation}
    \frac{1}{P} \frac{dP}{dz} = -\frac{z\,GM_*}{c_{s}^{2}(r^2 + z^2)^{3/2}}
\end{equation}

and can be solved through

\begin{equation}
    P(z) = c_{s}(z)^{2} \rho(z) = C \, \mathrm{exp} \left(\, - \int \frac{z\,GM_*}{c_{s}^{2}(r^2 + z^2)^{3/2}}  dz\, \right)
\end{equation}

which will obtain the vertical distribution for the gas density ($\rho(z)$). In DALI, the disk temperature structure is first obtained from the initial model, which computes the thermal conditions of the material considering all relevant cooling and heating processes, and is then used to update the gas density distribution accounting for thermal support. The updated density is consecutively used to recalculate the temperature for the next iteration and start over until reasonable convergence of the disk structure is achieved, which in this case is checked by verifying that the variation of the temperature and vertical location of the CO layer is minimal. The integration constant $C$ is adjusted such that $\int \rho(z)\,dz$ is preserved with each iteration. In the isothermal approximation $c_{s}^{2}$ does not have a vertical dependence and we recover the gaussian distribution indicated in equation 2.

Figure \ref{ttauri_model} shows the 2D maps for the gas density, temperature and CO abundance (with respect to H$_2$) of a converged DALI model for a 10$^{-2}$M$_{\odot}$ disk around a 1M$_{\odot}$ and 1L$_{\odot}$ stellar system. The static gaussian distribution corresponds to the initial DALI output, computed from the setup equations outlined in section 2.1. The self-consistent model solution is the result of six iterations solving the hydrostatic equilibrium equations, considering the previously computed thermal structure. The material appears to be more vertically extended in the self-consistent model, which is expected due to the increased pressure support of the heated atmosphere calculated from the vertical temperature gradient. Our study focuses on the $^{12}$CO $2-1$ emission surface, which is directly computed from the models as the location where the integrated column density of material results in an optical depth ($\tau$) of 1 for $^{12}$CO $2-1$. Indeed, the  $\tau =$1 layer of DALI models is consistent with the emission surfaces extracted from ALMA gas emission maps of  $^{12}$CO $2-1$ \citep{Paneque_2023_vert}.

\begin{figure*}[h!]
   \centering
   \includegraphics[width=\hsize]{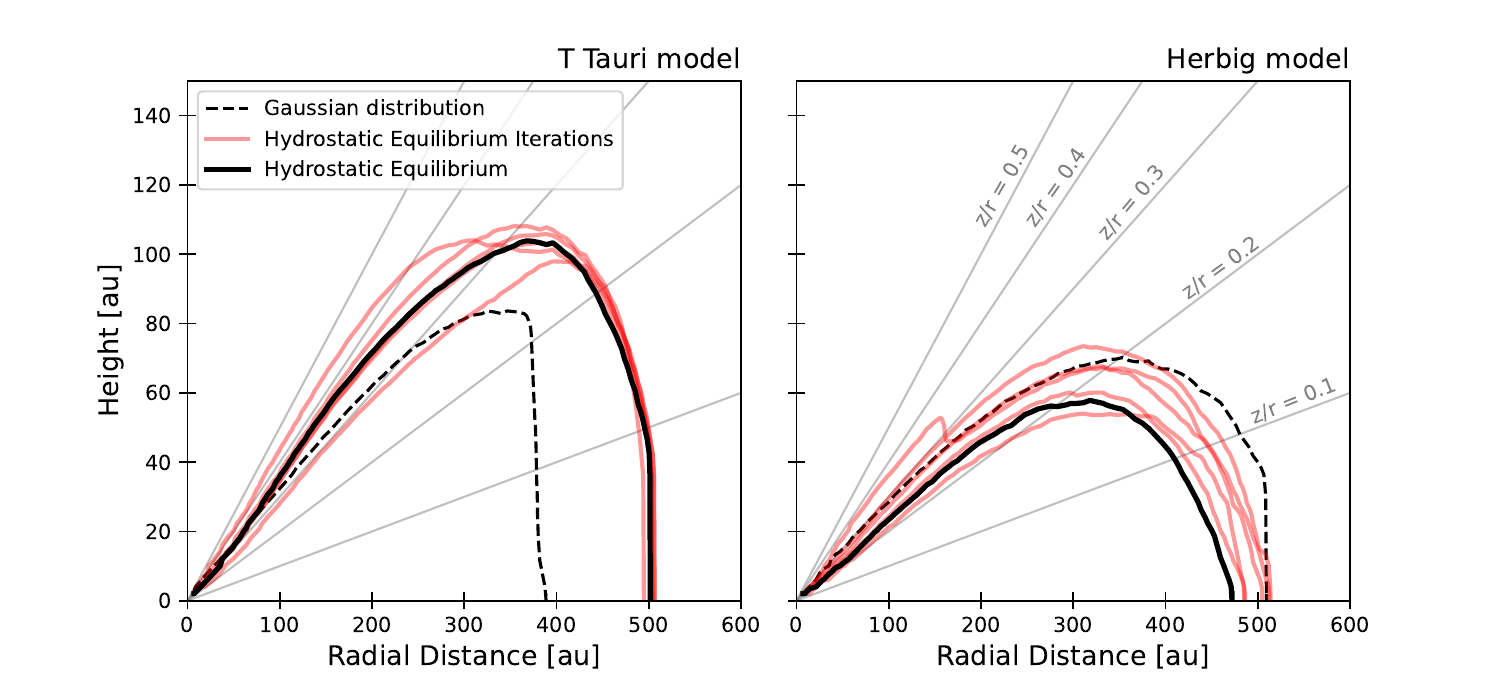}
      \caption{Variation of the $^{12}$CO $\tau = $ 1 surface when considering the hydrostatic equilibrium model solutions. Left panel shows the output for a T\,Tauri system and Right panel the results for a Herbig system. Both models correspond to a disk mass of 10$^{-2}$\,M$_{\odot}$. The dashed lines trace the surface from the parametric gaussian distribution and the solid black line the final hydrostatic equilibrium solution considered. Each red line represents a model iteration of hydrostatic equilibrium in DALI.
              }
         \label{hydro_models}
\end{figure*}

When accounting for hydrostatic equilibrium, the $\tau = 1$ surface varies both in its vertical and radial extent from the surface extracted from the parametric model. Figure \ref{hydro_models} indicated the variations for the $\tau = 1$ layer in a 10$^{-2}$\,M$_{\odot}$ disk around a 1\,M$_{\odot}$ and 1\,L$_{\odot}$ star (T\,Tauri model) compared to the same disk mass around a 2\,M$_{\odot}$ and 17\,L$_{\odot}$ star (Herbig model). As the hydrostatic equilibrium solution depends on the stellar mass, temperature and density distribution, it is expected that the surfaces will be different in each case. Convergence is reached rapidly after a couple of iterations, therefore all presented models in this work will consider the output from the 6th iteration of the self-consistent hydrostatic equilibrium solver. Small perturbations can be identified in the surfaces extracted from the first iterations, but we do not consider them relevant as they are artifacts that appear while the model is reaching convergence.

\section{Results}

As has been shown in multiple studies, CO molecular surfaces can be parametrized through an exponentially tapered power-law \citep{MAPS_Law_Surf, Law_2023_CO_isotop_surf, Paneque_2023_vert}. To model each $^{12}$CO $\tau = 1$ surface from the self-consistent DALI outputs after solving the hydrostatic equilibrium equations, we select the following prescription,

\begin{equation}
z(r) = z_0\times \left(\frac{r}{100\,\mathrm{au}}\right)^{\phi} \times \exp\left[\left(\frac{-r}{r_{\mathrm{taper}}}\right)^{\psi}\right]
\end{equation}

\begin{figure}[h!]
   \centering
   \includegraphics[width=\hsize]{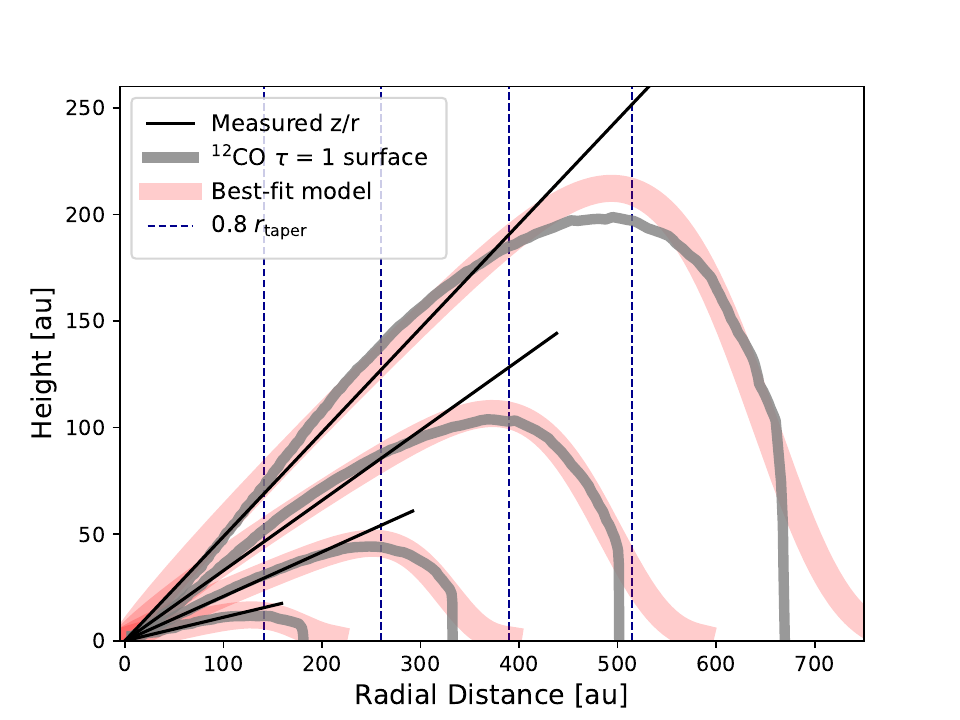}
      \caption{Schematic of the method use to measure the $^{12}$CO surface $z/r$ values. Each surface traced by the grey line is the output of a T\,Tauri model, in order of increasing height and radial extent they have disk masses of 10$^{-4}$ , 10$^{-3}$ , 10$^{-2}$ and 10$^{-1}$\,M$_{\odot}$. The shaded red curve shows the best-fit parametric model fitted to the molecular surface, from where $r_\mathrm{taper}$ is the taper radius. The measured $z/r$ is obtained by fitting the surface region within 80\% of $r_\mathrm{taper}$, this distance is indicated by the vertical dashed line. The characteristic $z/r$ is traced by the straight black line for each surface.
              }
         \label{method_zr}
\end{figure}

The best fit values for $z_0$, $\phi$, $r_{\mathrm{taper}}$ and $\psi$ are obtained using a Monte Carlo Markov Chain (MCMC) sampler
as implemented by emcee \citep{emcee_ref}. We initialize the MCMC
assuming a uniform distribution of the four parameters within the following intervals: $z_0 \in [0, 100]$, $\phi \in [0, 10]$,
$r_{\mathrm{taper}}/\mathrm{au} \in [50, 500]$, $\psi \in [0,10]$. We use 30 walkers, with 500 steps as burn-in and 800 steps to evaluate confidence intervals. Each surface is then identified based on its characteristic $z/r$ value, which corresponds to the best-fit line of constant $z/r$  that traces the height profile within 80\% of $r_{\mathrm{taper}}$ \citep{Law_2023_CO_isotop_surf}. This assures us that we are characterizing the rising portion of the $^{12}$CO surface, as exemplified in Figure \ref{method_zr} where a schematic of the method is shown. We consider only the rising portion to match the observational prescription found in the literature \citep{Law_2022_12CO, Law_2023_CO_isotop_surf} and to avoid the drop of the emission surface in the outer disk, an effect that happens due to the radially decreasing density profile. In some model outputs this drop appears as a sudden, almost vertical, feature in the outer radii which is likely a resolution artifact that does not impact our results.

\begin{figure*}[h!]
   \centering
   \includegraphics[width=\hsize]{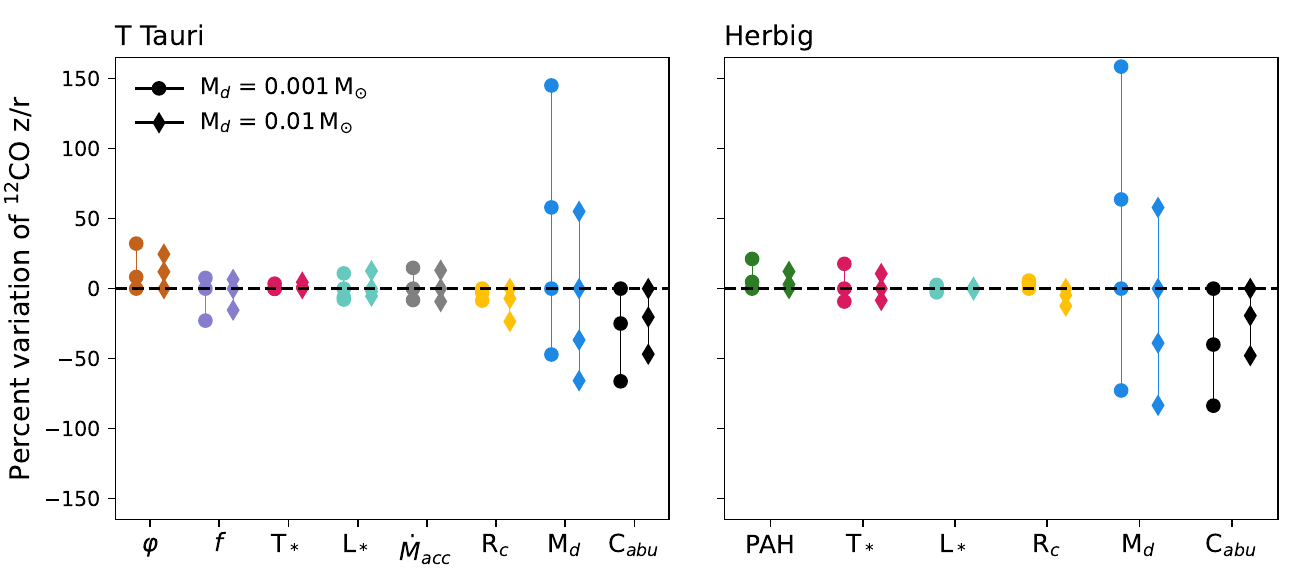}
      \caption{Percent variation of the $^{12}$CO surface $z/r$ caused by various stellar and disk parameters. Left panel indicates the results for the T\,Tauri system and right panel for the Herbig system. Each parameter corresponds to a specific color. The circle markers indicate the effects on a disk with mass 10$^{-3}$\,M$_{\odot}$ and the diamond markers the effects on a disk with mass 10$^{-2}$\,M$_{\odot}$. Specific details on the parameter range and properties are shown in Table \ref{table_DALI_varyparam}.
              }
         \label{vary_param_grid}
\end{figure*}

\begin{figure*}[h!]
   \centering
  \includegraphics[width=\hsize]{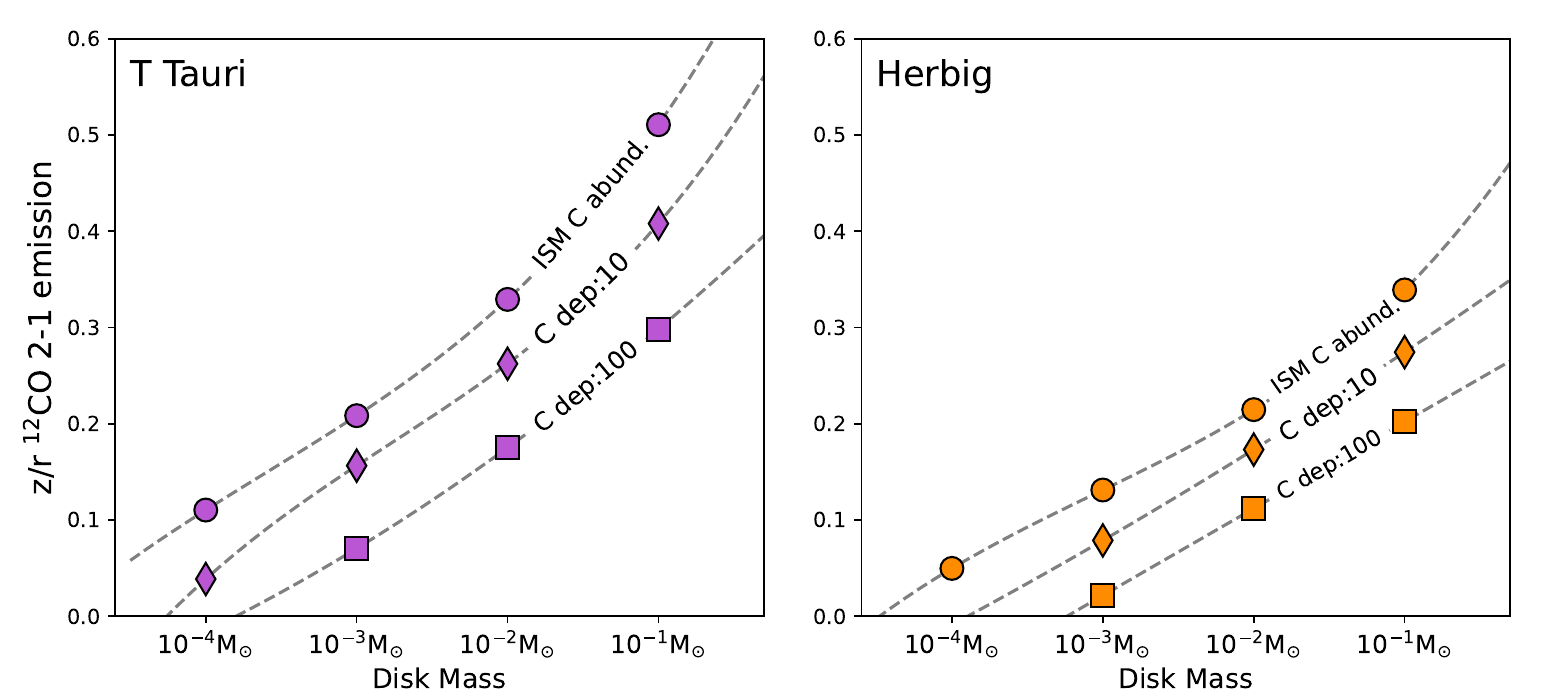}
      \caption{Relation between the $^{12}$CO surface $z/r$ and the total disk mass ($z/r$ - $M_d$ relationship) with a dependence on the volatile carbon abundance. Left panel shows results for T\,Tauri systems and right panel for Herbig systems.  Circles indicate ISM-like carbon abundance (no depletion), diamonds trace a carbon depletion of 10 and squares a depletion of 100, with respect to the ISM.
              }
         \label{CO_dep_zr_results}
\end{figure*}

\subsection{Parameters that affect vertical CO surface}

Our fiducial models for each stellar type are indicated in Table \ref{table_DALI_param}. To determine which parameter variation most strongly affects the measured $z/r$ we compute additional models with a varied stellar luminosities and temperatures as reported in Table \ref{table_DALI_varyparam} covering the range of reported values from the comparison observations. Disk structure values for the flaring angle ($\varphi$), small-to-large dust grain fraction ($f$), PAH abundance, surface density characteristic radius ($R_c$), total disk mass ($M_d$) and volatile carbon abundance (C$_{\mathrm{abu}}$) are also sampled. Only one parameter is varied at a time in each model, starting from the fiducial values. For every case the characteristic $z/r$ is computed as described at the beginning of Section 3. Figure \ref{vary_param_grid} presents the percentual variation from the fiducial model for disk masses of 10$^{-3}$\,M$_{\odot}$ and 10$^{-2}$\,M$_{\odot}$. The full vertical profiles for each parameter can be found in Figures \ref{vary_param_appendix_M1e-2} and \ref{vary_param_appendix_M1e-3}.

\begin{table*}[h!]
    \centering
    \def\arraystretch{1.5}
    \setlength{\tabcolsep}{6pt}
    \caption{Observational sample, stellar parameters and estimated total disk masses}
    \begin{tabular}{c|c|c|c|c|c|c|c|c}
    \hline
    \hline
         \multirow{2}{*}{Star}& M$_*$ & L$_*$ & T$_*$ &  \multicolumn{4}{c|}{Previous disk mass estimates [M$_{\odot}$]} & \multirow{2}{*}{References}\\
          & [M$_{\odot}$] & [L$_{\odot}$] & [K] & Dust mass $\times$ 100 & CO flux & HD emission & Dynamical & \\
    \hline
    \multicolumn{9}{l}{\textbf{T\,Tauri Disks}}\\
    \hline
    MY Lup & 1.27  & 0.87 & 5129 & 1.5$\times$10$^{-2}$ &  8.3$\times$10$^{-5}$ & - & - & 4 , 3, 12, 5\\
    \hline
    AS 209 & 1.2 & 1.41 &  4266 & 7.5$\times$10$^{-2}$ &  4.5$\times$10$^{-3}$ & -  & $>$2$\times$10$^{-4}$ & 1, 13, 16, 18\\
    \hline
    WaOph 6 &  1.12 & 2.9 & 4169 & 1.4$\times$10$^{-2}$ & - & - & - & 4, 3, 14 \\
    \hline
    IM Lup & 1.1 &  2.57 & 4266& 3.67$\times$10$^{-1}$ &  2$\times$10$^{-1}$ & - &1.1$\times$10$^{-1}$ & 1, 13, 16, 18\\
    \hline
    GM Aur & 1.1 & 1.2 &  4350& 7.3$\times$10$^{-2}$  &  2$\times$10$^{-1}$ & (2.5–20.4)$\times$10$^{-2}$ & 1.2$\times$10$^{-1}$ & 1, 13, 16, 18, 20\\
    \hline
     DoAr 25 & 1.06  & 0.95 & 4266 & 4.2$\times$10$^{-2}$ & - & - & - &4, 3, 5\\
    \hline
    LkCa 15 & 0.7  & 1.098 & 4162 & 2.7$\times$10$^{-2}$ & - & $<$6.2$\times$10$^{-2}$ & - & 5, 20\\
    \hline
    GW Lup &  0.62  & 0.33 & 3631 & 1.5$\times$10$^{-2}$ & 9.6$\times$10$^{-5}$ & - & -  &4, 3, 12, 5\\
    \hline
    Sz 91 &  0.55 & 0.26 & 3720 & 2.2$\times$10$^{-3}$ & - & - & - & 4, 10, 5\\
    \hline
    DM Tau &  0.50  & 0.24 & 3705 & 1.6$\times$10$^{-2}$ & - & (1.0–4.7)$\times$10$^{-2}$  & - & 4, 9, 5, 20\\
    \hline
    Elias 2-27 & 0.46  & 0.91 & 3890 & 3.4$\times$10$^{-2}$ & - & - & 8$\times$10$^{-2}$ & 2, 3, 5, 17\\
    \hline
    \hline
    \multicolumn{8}{l}{\textbf{Herbig Disks}} \\
    \hline
    HD 100546 &  2.10  & 23.4 & 9750 & 1.2$\times$10$^{-2}$ & 3.2$\times$10$^{-2}$ & $<$8$\times$10$^{-2}$ &- & 4, 8, 11, 15, 19\\
    \hline
    HD 97048 & 2.70  &  44.2 & 10500 & 4.7$\times$10$^{-2}$ & 1$\times$10$^{-1}$ & $<$9.4$\times$10$^{-2}$ & - & 4, 8, 11, 15, 19 \\
    \hline
    HD 100453 & 1.66  & 6.31 & 7400 & 5.3$\times$10$^{-3}$ & 3.2$\times$10$^{-2}$ & $<$1$\times$10$^{-2}$ & - & 6, 11, 15, 19 \\
    \hline
    HD 34282 &   1.59 & 13.64 &  8625 & 2.6$\times$10$^{-2}$ & 1.2$\times$10$^{-1}$ & - & - & 7, 11, 15 \\
    \hline
    MWC 480 & 2.1 & 21.9 & 8250 & 1.2$\times$10$^{-1}$ & 1$\times$10$^{-1}$ & -  & 1.5$\times$10$^{-1}$ & 1, 13, 15, 16\\
    \hline
    HD 163296 & 2.0 & 17.0 &  9332 & 8.4$\times$10$^{-2}$ & 1$\times$10$^{-1}$ & $<$6.7$\times$10$^{-2}$  &  1.3$\times$10$^{-1}$ & 1, 13, 15, 16, 19\\
    \hline
    HD 142666 & 1.73 & 9.1&  7500 & 7.6$\times$10$^{-3}$ & 3.2$\times$10$^{-2}$ & - & - & 4, 8, 11, 15 \\
    \hline
    Ak Sco &  1.2 & 2.4 & 6250 & 1.8$\times$10$^{-3}$ & 2.3$\times$10$^{-3}$ & - & - & 8, 11, 15 \\

    \hline
    \hline
    \end{tabular}

    \tablebib{ (1) \citealt{MAPS_Oberg} (2) \citealt{Veronesi_2021_Elias} (3) \citealt{DSHARP_Andrews} (4) \citealt{Law_2022_12CO} (5) \citealt{Manara_2023_ppvii} (6) \citealt{Rosotti_2020} (7) \citealt{Merin_2004} (8) \citealt{Fairlamb_2015}
    (9) \citealt{Pegues_2020} (10) \citealt{Tsukagoshi_2019} (11) \citealt{stapper_2022_dustmass} (12) \citealt{Miotello_2017} (13) \citealt{sierra_2021_MAPS} (14) \citealt{Brown-Sevilla_2021} (15) \citealt{stapper_2024_gasmass} (16) \citealt{Martire_2024_stratified_dyn_mass} (17) \citealt{Veronesi_2021_Elias} (18) \citealt{MAPS_Zhang} (19) \citealt{Kama_2020_HD} (20) \citealt{McClure_2016_HD}}

    \label{table_sample_all}
\end{table*}

The largest variation in $z/r$ is caused by varying the total disk mass, which is sampled at 10$^{-4}$\,M$_{\odot}$ , 10$^{-3}$\,M$_{\odot}$ , 10$^{-2}$\,M$_{\odot}$ and 10$^{-1}$\,M$_{\odot}$. An order of magnitude change in the disk mass can vary the emission height by more than 50\%. Volatile carbon abundance also presents a strong effect in the emission surfaces. Indeed, an order of magnitude decrease in the assumed volatile carbon abundance with respect to the ISM can lower the characteristic $z/r$ by $\sim$30\% and two orders of magnitude in carbon depletion produce $z/r$ variations beyond 50\%. These two parameters have a degenerate contribution to the vertical surface, as can be seen from Figure \ref{CO_dep_zr_results}. Increasing disk mass will push the CO emission layer higher, while for the same disk mass a lower volatile carbon abundance (higher depletion) will push the emitting layer towards the midplane. This result can be intuitively understood by remembering that the studied CO emision layers correspond to the location of the optically thick $\tau = 1$ surface and, therefore, they trace the vertical location at which the vertical column density of CO gas saturates, as seen from outside of the disk structure towards the midplane. A more massive disk will have larger column densities and saturate higher above the midplane. Likewise, for the same disk mass, a lower amount of CO gas due to lower quantities of volatile carbon will diminish the CO column densities and push the emission surface towards the midplane. This effect of carbon abundance in the vertical structure had been previously observed in \citet{Calahan_2021_HD_TwHya} when constraining the disk model of TW Hya.

Across all masses, for the same degree of carbon depletion, the T\,Tauri systems display a larger $z/r$ than the disks around Herbig stars (see Fig. \ref{CO_dep_zr_results}). This is expected due to the larger stellar mass of the Herbig model, which will translate into a higher gravitational potential (equation 5). The parameter space between our models is covered using a cubic spline which we use to extrapolate our results beyond our mass grid, to nearby values. We refer to this result as the $z/r$-$M_d$ relationship, noting that volatile carbon depletion will shift it and must be accounted for.

All other parameters, including order of magnitude variations in stellar accretion rate and PAH abundance (see Table \ref{table_DALI_varyparam}), cause a variation of less than $\sim$30\%. We stress that these results only hold for the $^{12}$CO $2-1$ surface $z/r$, other molecules, particularly those more sensitive to UV fields are likely more significantly affected by some of the sampled parameters.

\begin{figure*}[h!]
   \centering
   \includegraphics[width=\hsize]{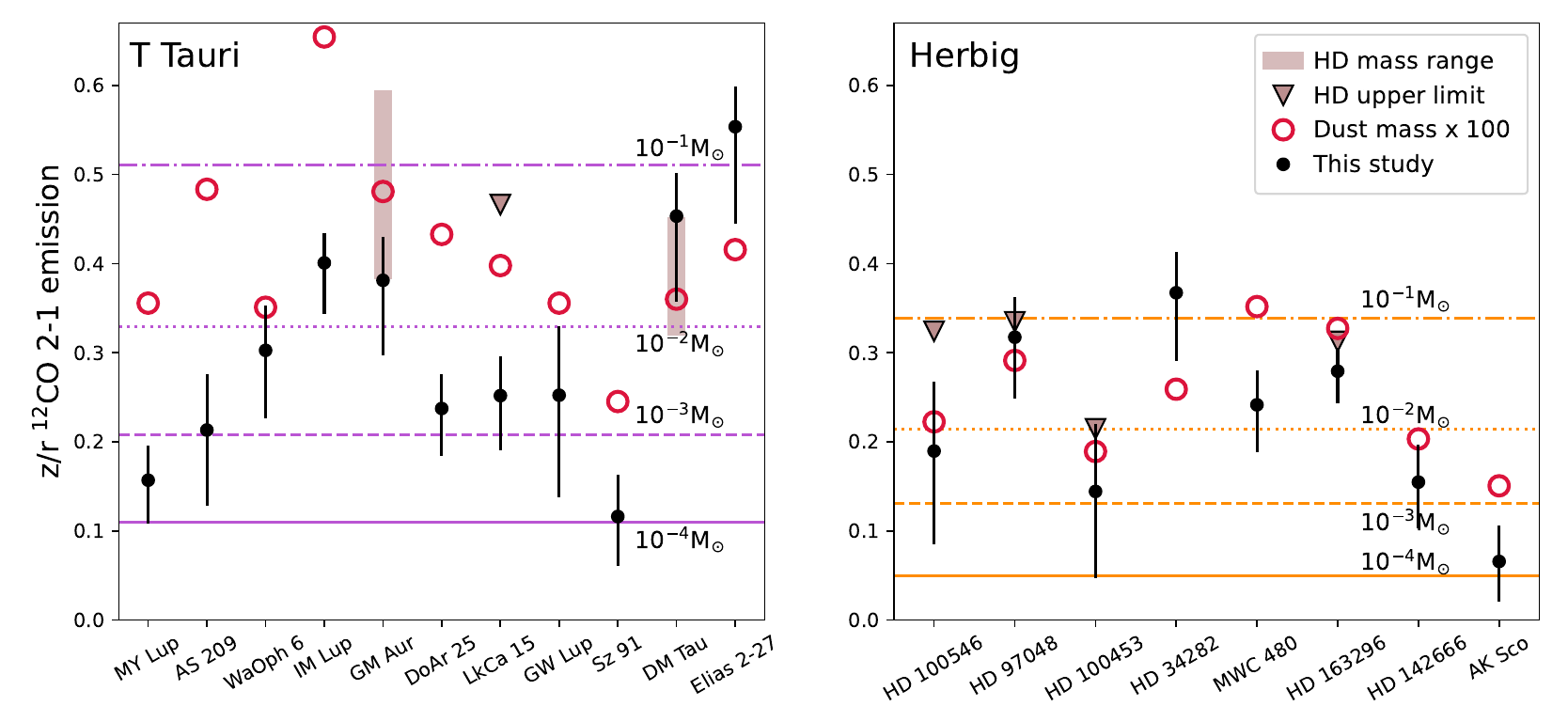}
      \caption{Comparison between the measured $^{12}$CO surface $z/r$ for a sample of disks and their expected values, based on literature total disk mass estimates from dust continuum observations. Black dots indicate the $z/r$ values obtained in this study for our disk sample, open red circles trace the expected $z/r$ values estimated from the dust continuum total disk masses and downwards triangles or shaded areas the expected upper limits or range values of $z/r$ from HD measurements, respectively. Left panel shows the results for a T\,Tauri system and right panel for a Herbig system.  Each horizontal line indicates the $z/r$ of our DALI models with disk masses of 10$^{-4}$\,M$_{\odot}$ , 10$^{-3}$\,M$_{\odot}$ , 10$^{-2}$\,M$_{\odot}$ and 10$^{-1}$\,M$_{\odot}$, in increasing $z/r$ order respectively.
              }
         \label{obs_comp_zr_COISM}
\end{figure*}

\subsection{Model comparison to observations}

In order to benchmark our model results and the $z/r$-$M_d$ relationship, we estimate the gas masses and expected $z/r$ in an observational sample of 19 disks, as indicated in Table \ref{table_sample_all}. The studied disks all have estimates of their total disk mass from dust continuum observations, additionally, some of them have disk mass estimates from modelling of CO, HD or N$_2$H$^+$ fluxes and dynamical analysis of CO rotation curves.

Most disk mass values from dust continuum emission are extracted by converting the total continuum flux into a dust mass, assuming a dust temperature and grain opacity and then using the canonical assumption of a gas-to-dust ratio of 100 \citep{Hildebrand_1983_dustmass, Ansdell_2016_Lupus, andrews_2013_diskmass, Barenfeld_2016_uppersco, pascucci_2016_diskmass, stapper_2022_dustmass}. Excluding the sources from the MAPS program \citep{MAPS_Oberg}, the total disk masses estimated from CO millimeter emission for the Herbig sample are calculated in \citet{stapper_2024_gasmass} based on an extensive grid of models, as is the case of MY\,Lup and GW\,Lup from the T\,Tauri sample \citep{Miotello_2017}. For the disks in the MAPS program \citep[AS\,209, IM\,Lup, GM\,Aur, MWC\,480, HD\,163296][]{MAPS_Oberg} the total disk mass calculated from dust continuum accounts for dust scattering processes and has a more accurate fitting of the thermal conditions \citep{sierra_2021_MAPS}. On the other hand, the CO disk masses come from detailed thermochemical modelling of \citet{MAPS_Zhang}. Estimations and upper limits of disk mass from HD emission are available for some sources and these were also calculated using thermochemical disk models to match the fluxes or non-detection threshold \citep{McClure_2016_HD, Kama_2020_HD}. Through the combination of N$_2$H$^{+}$ and C$^{18}$O emission, thermochemical models assuming a degree of disk ionization have computed the total disk mass and expected carbon depletion \citep{Trapman_2022_N2Hp}. Dynamical masses have been obtained for some disks, through the detection of super-Keplerian motion in the CO gas rotation curves from the outer radius of the disk rotation profile using isothermal \citep{Veronesi_2021_Elias} and thermally stratified \citep{Martire_2024_stratified_dyn_mass} vertical disk prescriptions.

This sample has been constructed from mid-inclination disks with available $^{12}$CO $2-1$ data cubes from which the surfaces have been extracted and published in previous works \citep{Law_2022_12CO,Law_2023_CO_isotop_surf, Stapper_2023_Herbig_surf, Paneque_2023_vert}. In all cases, the vertical structure from the observations is obtained following the methodology presented in \citet{Pinte_2018_method}, where the surfaces are constrained based on the emission maxima across the molecular line emission channel maps. However, some have used the automated code DISKSURF \citep{disksurf} and others have employed the manual masking approach of the code ALFAHOR \citep{Paneque_2023_vert}. All $^{12}$CO surfaces for the Herbig disk sample are taken from \citet{Stapper_2023_Herbig_surf}, where ALFAHOR  is used to extract the vertical profile. For the T\,Tauri disks, part of the sample (AS\,209, WaOph\,6, IM\,Lup, GM\,Aur, Elias\,2-27) is presented in \citet{Paneque_2023_vert}, also using ALFAHOR. The remainder of the T\,Tauri disks are published in \citet{Law_2023_CO_isotop_surf} using an initial version of DISKSURF that in some cases underestimated the vertical profiles due to contamination from the back side of the disk \citep[for a full discussion on this issue see][]{Paneque_2023_vert}. In order to do a uniform analysis, the disks previously studied using DISKSURF (MY\,Lup, DoAr\,25, LkCa\,15, GW\,Lup, Sz\,91, DM\,Tau) are re-analysed using ALFAHOR, yielding comparable results but with a lower scatter. The scatter corresponds to the spread in height values recovered from the channel map analysis within each radial bin. A larger scatter, which may be due to low signal-to-noise data, contamination from the back side or optical depth effects for less abundant molecules, will produce a broader vertical profile, as the dispersion will be larger within a given radial bin. All vertical profiles from the observational sample are presented in Figures \ref{ttauri_obs} and \ref{herbig_obs}, compared to the vertical profiles from the fiducial DALI models of varying disk mass with ISM-like carbon abundance.

As done for the DALI models, the vertical profiles of the observational sample are processed to obtain their characteristic $z/r$. To accurately account for the spread in the vertical profile, $z/r$ is calculated for the mean value and also for the lower and upper limits of the profiles, which indicate the dispersion within radial bins and will give the uncertainty on the characteristic $z/r$.

\begin{figure*}[h!]
   \centering
   \includegraphics[width=\hsize]{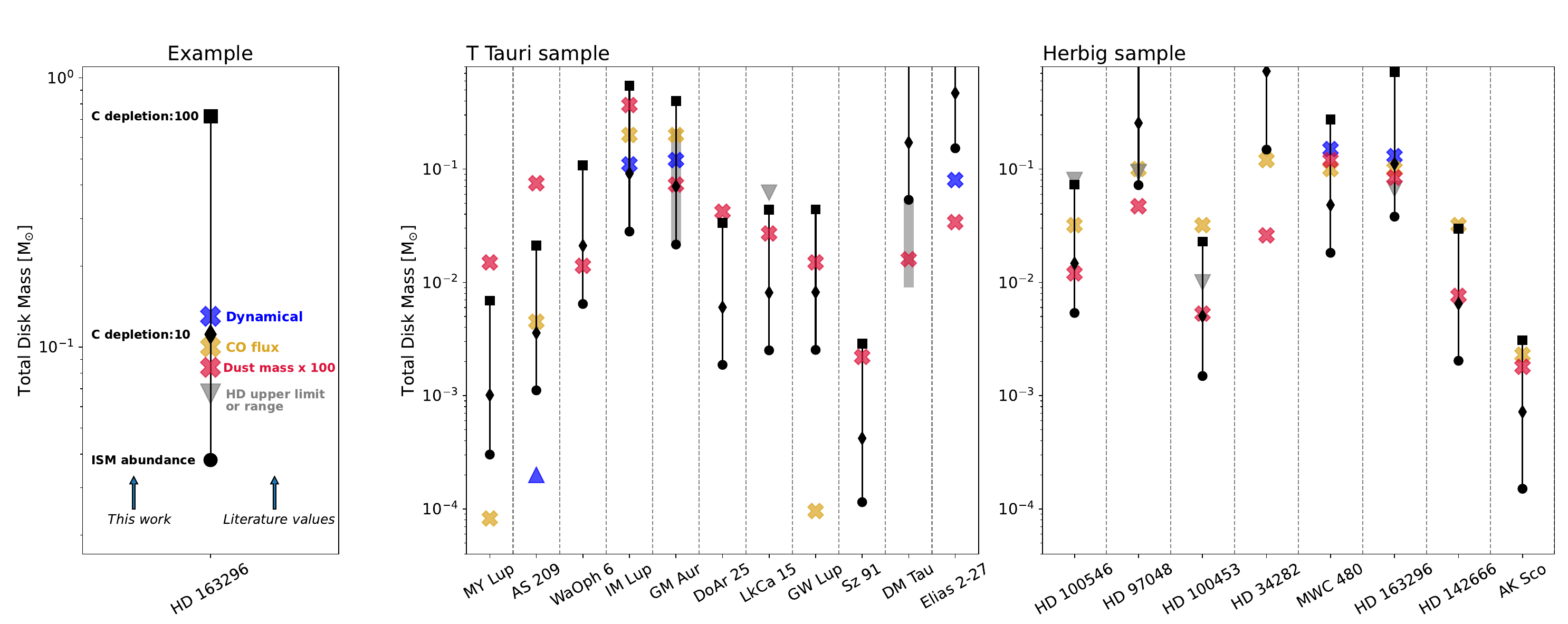}
      \caption{Estimated total disk masses from $z/r$ analysis, using varying carbon abundances, compared to literature values. Leftmost panel is an example and legend showing the Herbig system HD\,163296. Black symbols correspond to the estimates done in this work from the characteristic $z/r$, where circles, diamonds and squares have been calculated using the  $z/r$ - $M_d$ relationship curves for ISM-like, 10 and 100 values of volatile carbon depletion, respectively. For each disk the available literature values of the total disk mass are shown in colored crosses, depending on the methodology used for their calculation and grey triangles and shaded areas for the HD upper limits and mass range values, respectively. The center and rightmost panels show the results for the T\,Tauri and Herbig full samples.
              }
         \label{final_mass_plot}
\end{figure*}

Based on the $z/r$ - $M_d$ relationship shown in Figure \ref{CO_dep_zr_results} we estimate the expected $z/r$ for the reported total disk mass derived from dust continuum and available HD upper limits, assuming an ISM-like carbon abundance. These values are compared against the observationally derived $z/r$ values in Figure \ref{obs_comp_zr_COISM}. For the majority of the sample, if we assume an ISM-like carbon abundance, our measured $z/r$ is below the expected value based on the dust masses considering a gas-to-dust ratio of 100. The only exceptions are DM\,Tau, Elias\,2-27, HD\,97048 and HD\,34282. Throughout the sample, the differences are more noticeable in the T\,Tauri systems, as Herbig systems have better agreement within the uncertainty of our method. We note that taking into consideration carbon depletion would lower the expected $z/r$ of the literature mass measurements (see Fig. \ref{CO_dep_zr_results}), obtaining a better agreement with the observational emission heights.

Overall, these results may indicate that the assumed gas-to-dust ratio in the systems is much lower than 100 (effectively lowering the literature estimates from the dust flux) or that our model assumptions are failing to account for a significant factor affecting the $z/r$ value, such as the presence of carbon depletion. We favor the second scenario as dust-derived masses are expected to be lower limits of the total disk mass due to the assumptions on uncertain dust properties required for their calculation and usually unaccounted for effects such as optical depth and scattering \citep[see discussion and references in][]{Miotello_ppvii_2023}. Additionally, dynamical estimates of total disk mass have shown typical gas-to-dust ratios $\gtrsim$100 \citep{Lodato_2023_mass_imlup, Martire_2024_stratified_dyn_mass}, further indicating that the total disk mass derived from dust measurements should be a lower limit to the actual value.

\subsection{Accounting for volatile carbon depletion}

The initial abundance of volatile carbon in our models will determine the amount of CO formed in each case. We have shown in the previous sections that this has a significant effect in the extracted vertical CO gas surfaces, with low-carbon models producing shallower emission layers. There is independent observational motivation to assume carbon depletion for our disk sample, as measurements of CO abundance in some of the sources have shown depletion factors of 10-100 from the typically assumed ISM value of $\sim$10$^{-4}$ \citep{Zhang_2019_CO_abundance,MAPS_Zhang}.

Figure \ref{final_mass_plot} compares the published masses from various methods (see Table \ref{table_sample_all}) to the derived disk mass from our $z/r$ values, calculated accounting for varying degrees of carbon depletion in the $z/r$ - $M_d$ relationship. For disks that show a large scatter in their literature values we focus on the total disk mass extracted from the dust mass assuming a gas-to-dust ratio of 100 as our main indicator of the expected value, unless a dynamical mass estimate is available in which case we prefer that value.

In some disks it is sufficient to consider a volatile carbon depletion of 10 to match the dust or dynamical masses, however many require carbon depletion of 100 to be in agreement, particularly in the T\,Tauri systems. We briefly discuss here the case of HD\,163296 and MWC\,480, which have well constrained CO depletion profiles from optically thin C$^{18}$O and C$^{17}$O emission \citep{MAPS_Zhang}. The HD\,163296 literature analysis indicates an almost constant CO abundance profile, depleted by a factor 10 with respect to the ISM value beyond $\sim$50\,au and  MWC\,480 follows a similar pattern going towards even lower values of CO depletion beyond $\sim$100\,au \citep{MAPS_Zhang}. Our analysis indicates that for HD\,163296 a global volatile carbon depletion of $\sim$10 is required to match the literature disk mass estimates to those derived from the CO $z/r$. For MWC\,480 the preferred carbon depletion from our results ranges between 10-100, also in agreement with the spatially resolved values of \citet{MAPS_Zhang}.

For all disks, the inferred total mass we obtain from the characteristic $z/r$, for each carbon depletion value, are presented in Table \ref{table_results_mass_CO} together with the preferred carbon depletion. For the most massive disks, assuming carbon depletion of 100 produces unrealistic disk mass values of $\gtrsim1$M$_{\odot}$ as the extrapolation goes largely beyond our parameter space grid and therefore we do not include these estimations in Table \ref{table_results_mass_CO}. The volatile carbon depletion value should be taken with caution, as it is a single order of magnitude approximation based on full-disk model predictions compared to the available dynamical or dust-derived total disk mass. In the optimal scenario it may be a lower limit of the disk integrated volatile carbon depletion with respect to the ISM.

\begin{table*}[h!]
    \centering
    \def\arraystretch{1.5}
    \setlength{\tabcolsep}{6pt}
    \caption{Studied systems, their calculated characteristic $^{12}$CO $z/r$ and associated total disk mass for various volatile carbon depletion scenarios.}
    \begin{tabular}{|c|c|c|c|c|c|}
    \hline
    \hline
             \multirow{2}{*}{Star} & Characteristic& \multicolumn{3}{c|}{Total disk mass [log$_{\mathrm{10}}$(M$_{\odot}$)]} & Preferred\\
          & $^{12}$CO $z/r$ & ISM C abundance & C depletion: 10 & C depletion: 100 & carbon depletion  \\
    \hline
    \multicolumn{6}{l}{\textbf{T\,Tauri Disks}}\\
    \hline
    MY Lup & 0.16$_{-0.05}^{+0.03}$ & -3.52$_{-0.48}^{+0.38}$ & -2.99$_{-0.44}^{+0.37}$ & -2.16$_{-0.44}^{+0.33}$ & 100\\
    \hline
    AS 209 & 0.21$_{-0.08}^{+0.05}$ & -2.95$_{-0.85}^{+0.53}$ & -2.45$_{-0.81}^{+0.55}$ & -1.67$_{-0.74}^{+0.49}$ & 100 \\
    \hline
    WaOph 6 & 0.30$_{-0.07}^{+0.05}$ & -2.19$_{-0.63}^{+0.34}$ & -1.67$_{-0.64}^{+0.35}$ & -0.97$_{-0.59}^{+0.36}$ & 10\\
    \hline
    IM Lup & 0.40$_{-0.05}^{+0.03}$ & -1.55$_{-0.34}^{+0.17}$ & -1.02$_{-0.35}^{+0.18}$ & -0.26$_{-0.39}^{+0.21}$ & 10\\
    \hline
    GM Aur & 0.38$_{-0.08}^{+0.05}$ & -1.67$_{-0.56}^{+0.27}$ & -1.13$_{-0.57}^{+0.27}$ & -0.40$_{-0.6}^{+0.32}$ & 10-100\\
    \hline
     DoAr 25 & 0.24$_{-0.05}^{+0.03}$ & -2.73$_{-0.50}^{+0.32}$ & -2.22$_{-0.49}^{+0.33}$ & -1.48$_{-0.43}^{+0.30}$ & 100 \\
    \hline
    LkCa 15 & 0.25$_{-0.06}^{+0.04}$ & -2.60$_{-0.56}^{+0.34}$ & -2.09$_{-0.56}^{+0.35}$ & -1.36$_{-0.49}^{+0.33}$ & 100\\
    \hline
    GW Lup & 0.25$_{-0.11}^{+0.08}$ & -2.60$_{-1.1}^{+0.59}$ & -2.09$_{-1.08}^{+0.6}$ & -1.36$_{-0.96}^{+0.58}$ & 10-100\\
    \hline
    Sz 91 & 0.12$_{-0.06}^{+0.04}$ & -3.94$_{-0.52}^{+0.47}$ & -3.38$_{-0.46}^{+0.43}$ & -2.54$_{-0.55}^{+0.42}$ & 100\\
    \hline
    DM Tau & 0.45$_{-0.09}^{+0.05}$ & -1.27$_{-0.53}^{+0.22}$ & -0.73$_{-0.54}^{+0.23}$ & - & 0\\
    \hline
    Elias 2-27 & 0.55$_{-0.09}^{+0.05}$ & -0.82$_{-0.49}^{+0.17}$ & -0.27$_{-0.50}^{+0.17}$ & - & 0\\
    \hline
    \hline
    \multicolumn{6}{l}{\textbf{Herbig Disks}} \\
    \hline
    HD 100546 & 0.19$_{-0.1}^{+0.08}$ & -2.27$_{-1.31}^{+0.74}$ & -1.83$_{-1.09}^{+0.76}$ & -1.14$_{-1.15}^{+0.84}$ & 10 \\
    \hline
    HD 97048 & 0.32$_{-0.07}^{+0.04}$ & -1.14$_{-0.52}^{+0.27}$ & -0.57$_{-0.66}^{+0.41}$ & - & 0\\
    \hline
    HD 100453 & 0.14$_{-0.09}^{+0.08}$ & -2.83$_{-1.18}^{+0.87}$ & -2.30$_{-1.03}^{+0.76}$ & -1.64$_{-1.06}^{+0.83}$ & 10\\
    \hline
    HD 34282 & 0.37$_{-0.08}^{+0.04}$ & -0.83$_{-0.49}^{+0.24}$ &  -0.14$_{-0.69}^{+0.14}$ & - & 0 \\
    \hline
    MWC 480 & 0.24$_{-0.05}^{+0.04}$ & -1.74$_{-0.53}^{+0.32}$ & -1.31$_{-0.52}^{+0.37}$ & -0.56$_{-0.57}^{+0.41}$ & 10-100 \\
    \hline
    HD 163296 & 0.28$_{-0.04}^{+0.02}$ & -1.42$_{-0.29}^{+0.18}$ & -0.94$_{-0.34}^{+0.23}$ & -0.14$_{-0.39}^{+0.14}$ & 10 \\
    \hline
    HD 142666 & 0.15$_{-0.05}^{+0.05}$ & -2.69$_{-0.65}^{+0.48}$ & -2.19$_{-0.53}^{+0.42}$ & -1.52$_{-0.56}^{+0.45}$ & 10 \\
    \hline
    AK Sco & 0.07$_{-0.05}^{+0.03}$ & -3.82$_{-0.46}^{+0.48}$ & -3.14$_{-0.49}^{+0.43}$ & -2.51$_{-0.48}^{+0.43}$ & 100 \\

    \hline
    \hline
    \end{tabular}
    \label{table_results_mass_CO}
\end{table*}

\section{Discussion}

\subsection{Low gas-to-dust or high carbon depletion?}

It has been previously discussed in the literature that current CO gas models compared to observations do not allow us to distinguish between a low gas-to-dust system (lower gas mass than dust-derived total disk mass) or a high volatile carbon depletion scenario \citep{Miotello_2017, Calahan_2021_HD_TwHya}. Our results are in agreement with this entanglement. However we favor the high carbon depletion over low gas-to-dust due to evidence for CO abundance evolution over time in young stars \citep{Bergner_2020_CO, Zhang_2020_CO}, measurements of CO depletion in sources from our sample \citep{Zhang_2019_CO_abundance, MAPS_Zhang} and results from dynamical mass estimates that indicate gas-to-dust ratios $\gtrsim$100 \citep{Martire_2024_stratified_dyn_mass} rather than much less than 100.

As discussed in \citet{Miotello_ppvii_2023}, various methods can yield orders of magnitude differences in the calculated disk mass. This is particularly apparent for the literature total disk mass values of T\,Tauri systems MY\,Lup and GW\,Lup. For these disks our analysis indicates that a high carbon depletion of $\sim$100 is needed to reach the reported dust-derived mass value, but the reported gas mass, which is several ($\sim$2) orders of magnitude lower in both cases, was extracted without accounting for carbon depletion \citep{Miotello_2017}. It is therefore likely that a higher carbon depletion rate would put all values in agreement, an alternative that was suggested by \citet{Miotello_2017} to account for the low gas-to-dust ratios that the measurement would otherwise entail.

We note that for the Herbig sample, the gas mass estimates used in this work were mostly obtained from DALI models of isotopologue CO emission \citep{stapper_2024_gasmass}, measuring the fluxes and disk radii in these tracers. These models assumed an ISM volatile carbon abundance, with a parameter space similar to our models. However they obtain mass values higher than the dust-based masses, consistent with a gas-to-dust ratio $\gtrsim$100 (see right panel of Figure \ref{final_mass_plot}). We are confident that carbon depletion must happen at least for MWC\,480 and HD\,163296, as it has been measured independently through dedicated models and observations \citep{MAPS_Zhang}. HD\,100546 also has estimates of volatile carbon depletion between 2-10, consistent with our results \citep{Bruderer_2012, Kama_2016_HD100546}. The overall discrepancy between our work and analysis using models of CO isotopologue fluxes is the consideration of hydrostatic equilibrium in our models. We show in Figure \ref{lucas_comp_fluxes} that for high disk masses, when the gas emission becomes optically thick, not accounting for hydrostatic equilibrium can underestimate the CO flux, therefore overestimating the calculated disk masses. Due to the optical depth, the use of model fluxes in the high disk mass regime also leads to large uncertainties of up to an order in magnitude in the calculated mass, which also affects the comparison \citep{Miotello_2017, stapper_2024_gasmass}. Overall, this highlights the sensitivity of vertical surfaces as a probe for total disk mass and volatile carbon depletion.

While our results are in agreement with most literature works, showing that carbon depletion is present by 1-2 orders of magnitude with respect to ISM levels (see Table \ref{table_results_mass_CO} and previous references), some recent articles have argued that carbon depletion may not be as common in protoplanetary disk systems \citep{Ruaud_2022_CO, Pascucci_2023_CO}. The differences between these results and the rest may come from the assumed temperature structure and the effect of considering or not hydrostatic equilibrium \citep{Bosman_2018_CO, Ruaud_2022_CO}, which our models account for, similar to those of \citet{Ruaud_2022_CO}. Another fundamental difference is the semantics when referring to carbon or CO depletion. In this work, we define carbon depletion as the lack of volatile carbon with respect to ISM levels, which causes a depletion of the CO gas abundance. However, we are agnostic as to what specific mechanism is responsible for depleting the gas phase CO, beyond photodissociation and freeze-out which are included in the DALI standard network \citep{Bruderer_DALI_2013}. The models used in \citet{Ruaud_2022_CO} also model photodissociation and freeze-out, but additionally consider the conversion of CO ice to other species, which implies a further reduction in both gas and ice CO abundances from ISM levels, therefore requiring lower levels of depletion.

\subsection{Synergy with other methods and comparison to previous studies}

Using our $z/r-$$M_d$ relationships can be useful to have an initial idea on the system properties. Contrary to other methods, we are able to utilize a bright and readily detected molecule to have a first-order joint constraint of disk mass and carbon abundance. Indeed, if the calculated $z/r$ is $\geq0.3$ then the total disk mass is $\geq10^{-2}$M$_{\odot}$. For lower $z/r$, the uncertainty on the carbon abundance creates a span of several orders of magnitude in the total disk mass value. However, if gas-to-dust ratio of at least 100 is assumed, which is reasonable from accretion rate values that point towards gas-rich disks \citep{Manara_2016_acc, Manara_2023_ppvii} and dynamical mass estimates \citep{Lodato_2019_H_R,Martire_2024_stratified_dyn_mass}, this work has shown that the total mass calculated from the dust continuum flux may be used as an anchor to estimate the lower limit of carbon depletion (see section 3.3).

Combining the information from CO surfaces with dynamical mass estimates, as has been done for a small sample in this work would put strong constraints on the overall carbon depletion during planet formation. Furthermore, the use of rarer CO isotopologues and C-abundance-sensitive tracers such as N$_2$H$^{+}$ has demonstrated to be a good tool to discern between the effects of varying disk mass and volatile carbon depletion \citep{Anderson_2019_N2Hp, Anderson_2022_LupusMass, Trapman_2022_N2Hp}. N$_2$H$^{+}$ models have the uncertainty of the disk ionization rate and complication of relying on a fainter molecular line. However, in combination with measurements of the CO vertical profile it could be possible to determine all three parameters, key for understanding protoplanetary disk conditions. We highlight that our results for DM\,Tau and GM\,Aur using the CO surface and dust-derived disk masses are in great agreement with the total disk mass and carbon depletion values determined in \citet{Trapman_2022_N2Hp} through N$_2$H$^{+}$ and C$^{18}$O flux analysis and in \citet{McClure_2016_HD} for the HD flux study of DM\,Tau.

\begin{figure*}[h!]
   \centering
   \includegraphics[width=\hsize]{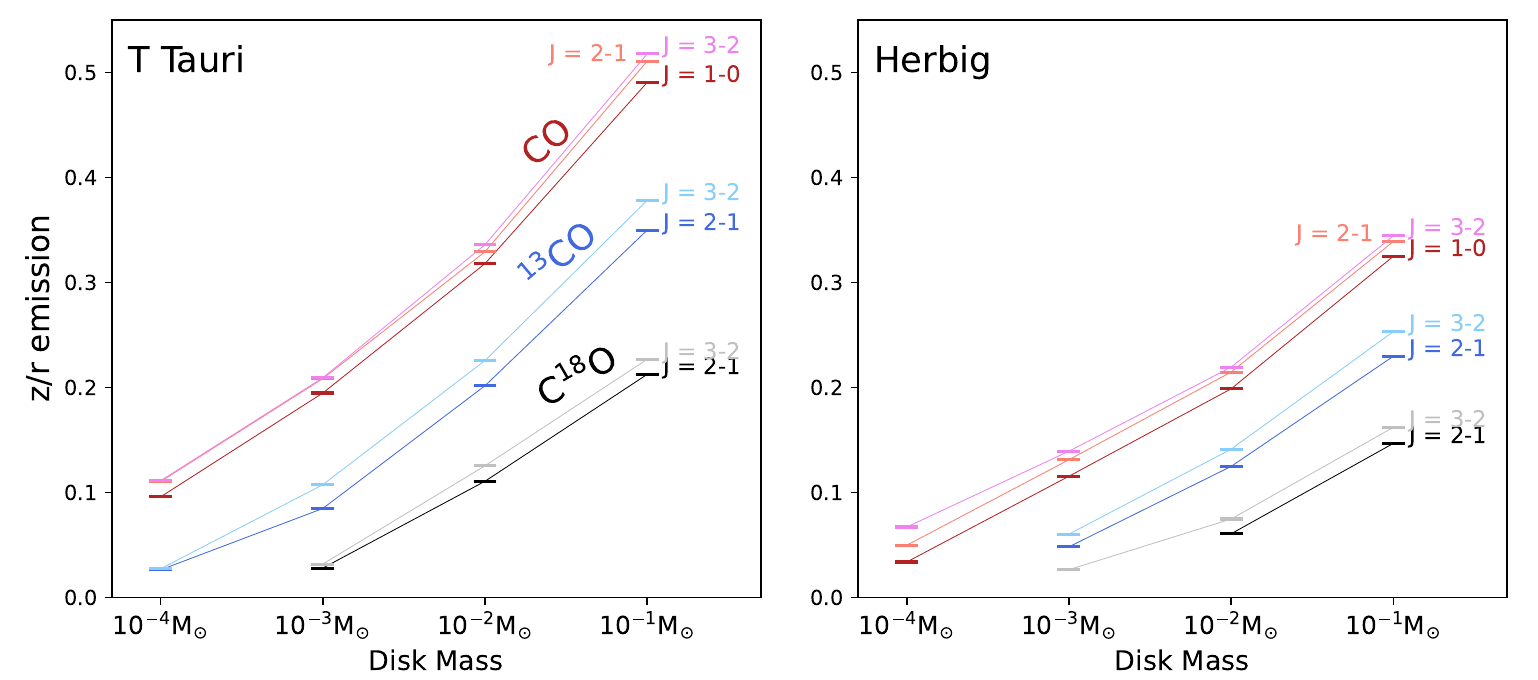}
      \caption{ Values of the $z/r - M_d$ relationship for various CO isotopes and transitions, assuming an ISM-like volatile carbon abundance. Left panel indicates results from the fiducial T\,Tauri models and the Right panel the same but for the Herbig stellar parameters case. Some low mass disk models are missing $z/r$ values for $^{13}$CO and/or C$^{18}$O due to their low abundance and lack of optically thick emission surface.
              }
         \label{CO_isotope_test}
\end{figure*}

Our models may also be used to interpret the observational $z/r - R_{\mathrm{CO}}$ trend, which predicts a relation between the location of the CO emission layer and the CO emission radius of the disk such that more elevated disks also have a larger radial extent \citep{Law_2022_12CO, Law_2023_CO_isotop_surf}. Thermochemical models have shown that the radial CO extent is directly related to the total disk mass \citep{Trapman_2024_diskmasssize} and our work finds that $z/r$ also scales with increasing disk mass. Therefore, the observational $z/r - R_{\mathrm{CO}}$ trend reflects on the amount of material contained in the disk. Both vertical and radial extent also depend on carbon abundance, therefore, in a strict interpretation, the observational trend offers information on the total CO gas mass, which can be translated to total disk mass if the carbon abundance is known.

\subsection{Caveats and future perspectives}

While exciting and very useful for an initial overview of the disk conditions, there are a few caveats to our results. In particular, the studied models have been developed using a full-disk prescription, a standard radial density profile and a single disk-integrated volatile carbon depletion value. However observations show disks with a plethora of substructure \citep[e.g.][]{DSHARP_Andrews,Andrews_2020_review, Long_2019_Taurus} and that CO abundance profiles varies radially \citep{Zhang_2019_CO_abundance, MAPS_Zhang}. Highly structured or transition disks, which is the case for several disks in our sample, are likely not accurately represented by our models. Future work, beyond the scope of this paper, will test the effect of large cavities and other substructure on the vertical profile and $z/r$-$M_d$ relationship.

We note that emission surfaces obtained at high enough spatial resolution may give direct information on the surface density profile, allowing for a more accurate description of the material distribution without the need to use a characteristic $z/r$ \citep[see for example,][]{Paneque_2023_vert}. Through this approach, the total disk mass may be more accurately determined and scaled depending on the carbon abundance in a source-by-source analysis, without the uncertainties of general models.

Most importantly, future work must focus on different molecular tracers, such as rarer CO isotopologues and bright optically thin molecules, for example CN and C$_2$H, that have been proposed to be sensitive tracers of the upper disk regions \citep{Cazzoletti_2018_CN, Bergin_2016_c2h, Paneque-Carreno_Elias1, paneque-carreno_2024_IMLup}. As a first step towards the analysis of other CO isotopologues and $^{12}$CO transitions, Figure \ref{CO_isotope_test} shows that for ISM-like carbon abundance the vertical layer of $^{13}$CO and C$^{18}$O will be more settled towards the midplane. The challenge in observing less abundant isotopologues is that it is only possible to retrieve optically thick emission surfaces for higher disk masses. An interesting take-away point from Figure \ref{CO_isotope_test} is that within the same isotopologue, different transitions create a negligible variation in the recovered $z/r$, as the emission comes from similar regions. This result is in agreement with previous theoretical studies and models of the vertical distribution of CO in disks \citep{van_zadelhoff_2003}. Our results in this work are only applicable to the CO emitting layer. However, it is likely that tracing the vertical structure with other molecular species will give direct insight into varied ongoing disk processes.

\section{Conclusions}

In this work we have presented a $z/r$-$M_d$ relationship, linking the vertical extent of $^{12}$CO $2-1$ to the disk mass and volatile carbon abundance. This was done through the study of thermochemical DALI models, computed accounting for hydrostatic equilibrium. Our models tested the effect of various stellar and disk parameters on the vertical structure and our theoretical predictions were benchmarked on a set of observations. The main findings and conclusions are the following:
\begin{enumerate}
    \item We determine that for T\,Tauri and Herbig systems, the total disk mass and volatile carbon abundance are the main parameters that set the location of the $^{12}$CO $2-1$ emission surface. This surface can be characterized by a constant characteristic $z/r$ tracing the rising portion of the vertical profile.
    \item The $z/r$-$M_d$ relationship for T\,Tauri systems is different than that of disks around Herbig stars. For the same disk mass and assumed volatile carbon abundance, material around T\,Tauri stars extends further vertically, as expected due to the difference in the stellar masses and gravitational potential.
    \item Using the total disk masses inferred from continuum dust observations (assuming a gas-to-dust ratio of 100) we are able to calibrate the $z/r$-$M_d$ relationship for each individual system and estimate the order of magnitude of carbon depletion.
    \item Fifteen out of the nineteen studied systems (79\%) show indicators of volatile carbon depletion (compared to the ISM abundance). For disks with previous constraints on their carbon abundance, our results yield comparable values. T\,Tauri disks seem to be more carbon depleted than Herbig systems.

\end{enumerate}

Overall, this study has demonstrated the utility of a bright and readily-detected molecular tracer, $^{12}$CO $2-1$, that may be used to probe two of the most fundamental disk parameters through its vertical extent: total disk mass and carbon abundance. As the sample of disks with high resolution data grows, studies of vertical layering in mid-inclination disks will be used to deepen our understanding not only on the thermal and density conditions, but also on the amount and composition of the materials available for planet formation.

\bibliographystyle{aa}
\bibliography{model_paper.bib}

\begin{acknowledgements}
This paper makes use of the following ALMA data:
\#2015.1.00192.S, \#2015.1.00168.S, \#2016.1.00344.S, \#2016.1.00204.S, \#2016.1.00484.L,  \#2016.1.00606.S, \#2017.1.00069.S and \#2018.1.01055.L ALMA is a partnership of ESO (representing its member states), NSF (USA), and NINS (Japan), together with NRC (Canada),  NSC and ASIAA (Taiwan), and KASI (Republic of Korea), in cooperation with the Republic of Chile. The Joint ALMA Observatory is operated by ESO, AUI/NRAO, and NAOJ.
Astrochemistry in Leiden is supported by the Netherlands Research School for Astronomy (NOVA), and by funding from the European Research Council (ERC) under the European Union’s Horizon 2020 research and innovation programme (grant agreement No. 101019751 MOLDISK).
GPR acknowledges support from the European Union (ERC Starting Grant DiscEvol, project number 101039651) and from Fondazione Cariplo, grant No. 2022-1217. Views and opinions expressed are, however, those of the author(s) only and do not necessarily reflect those of the European Union or the European Research Council. Neither the European Union nor the granting authority can be held responsible for them.

\end{acknowledgements}

\begin{appendix}

\section{Full emission surfaces from models and observational sample}

\begin{figure*}[h!]
   \centering
   \includegraphics[width=\hsize]{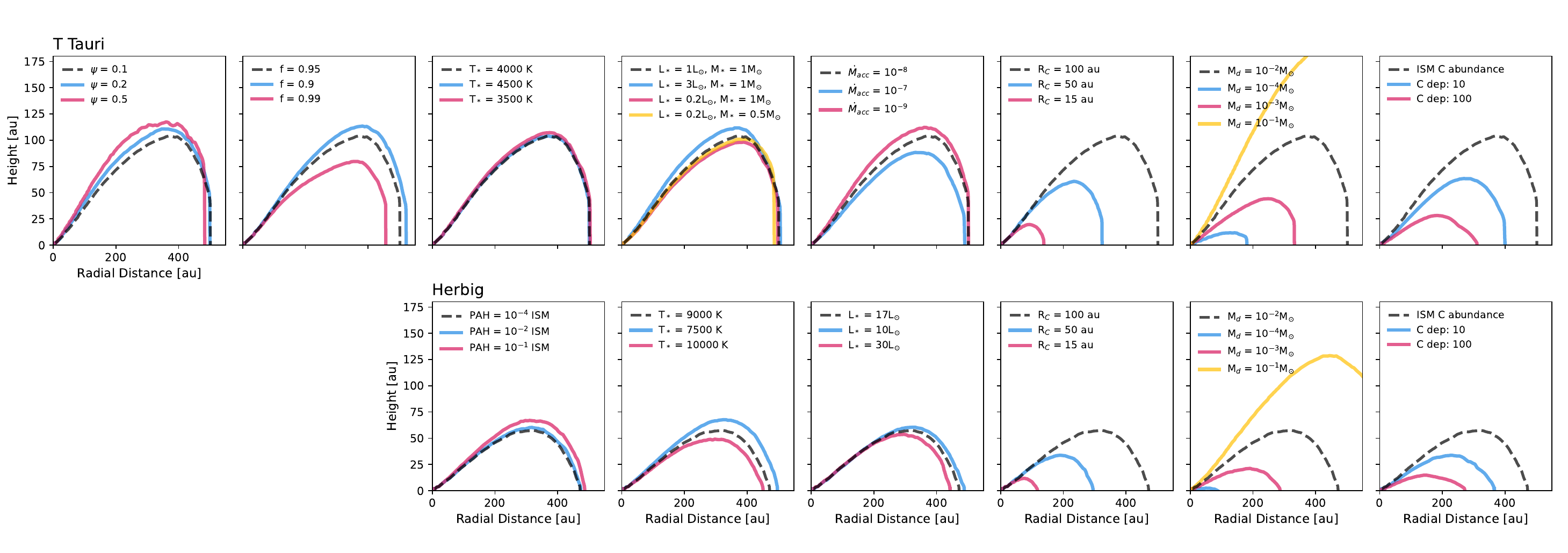}
      \caption{Vertical profiles tracing $\tau=1$ layer of the DALI model grid for $M_{\mathrm{disk}} = $10$^{-2}$\,M$_{\odot}$. Top row shows the sampled parameters for the T\,Tauri systems and bottom row for Herbigs. In each panel the black dashed line corresponds to the fiducial model and only one parameter is varied, as indicated by the legend and colors.
              }
         \label{vary_param_appendix_M1e-2}
\end{figure*}

\begin{figure*}[h!]
   \centering
   \includegraphics[width=\hsize]{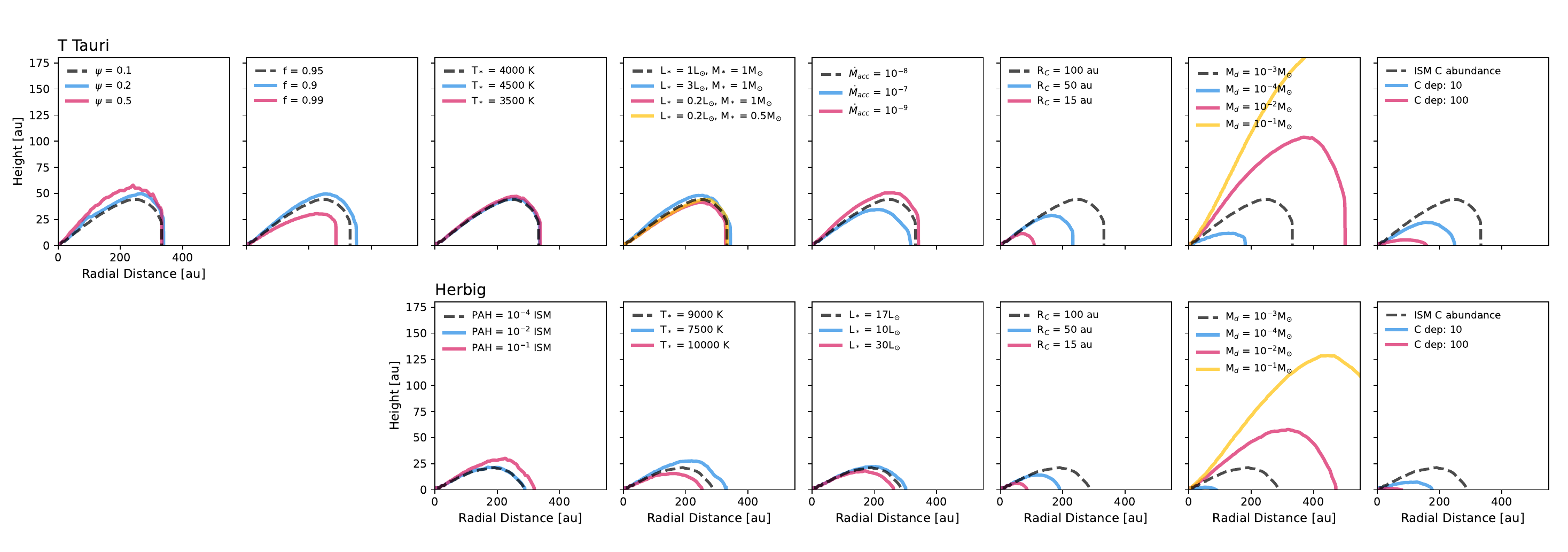}
      \caption{Same as Figure \ref{vary_param_appendix_M1e-2} but for fiducial model with $M_{\mathrm{disk}} = $10$^{-3}$\,M$_{\odot}$.
              }
         \label{vary_param_appendix_M1e-3}
\end{figure*}

In the main text our results are expressed in $z/r$. For visual comparison, here we show the full vertical profiles extracted for the CO emission surfaces in models and observations. Figures \ref{vary_param_appendix_M1e-2} and \ref{vary_param_appendix_M1e-3} show the emission surfaces tracing the $\tau=1$ layer, as extracted from the DALI model output of the fiducial models and models varying specific parameters (see Table \ref{table_DALI_varyparam} for details).

\begin{figure*}[h!]
   \centering
   \includegraphics[width=\hsize]{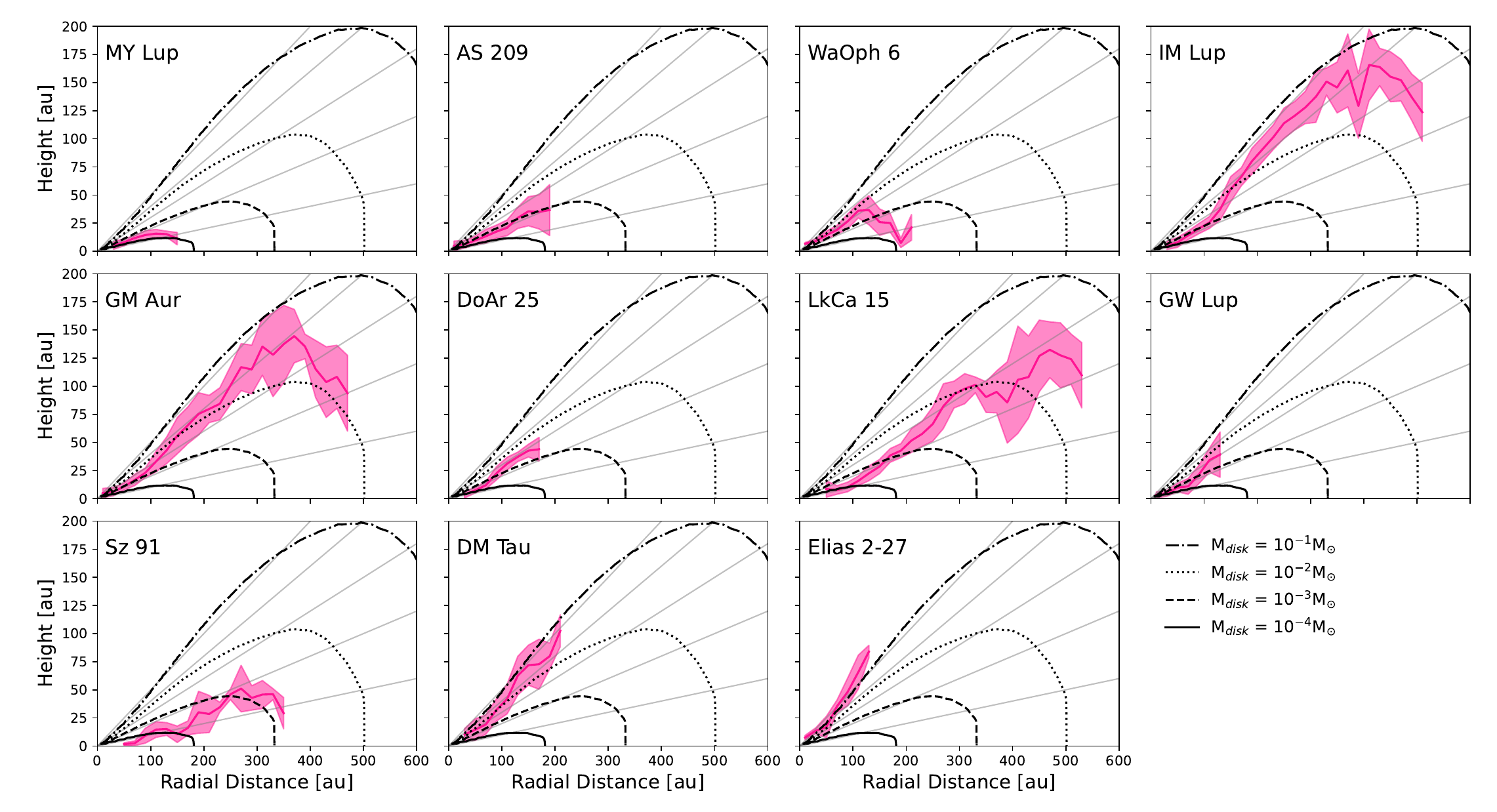}
      \caption{Observational CO emission surfaces for the T\,Tauri disk sample. Magenta lines indicate the observational results and data vertical dispersion, black lines show the $\tau=1$ layer from the fiducial DALI models using $R_c = 100$\,au for different disk masses indicated by different line styles. Grey lines mark constant $z/r$ surfaces of 0.1, 0.2, 0.3, 0.4 and 0.5.
              }
         \label{ttauri_obs}
\end{figure*}

\begin{figure*}[h!]
   \centering
   \includegraphics[width=\hsize]{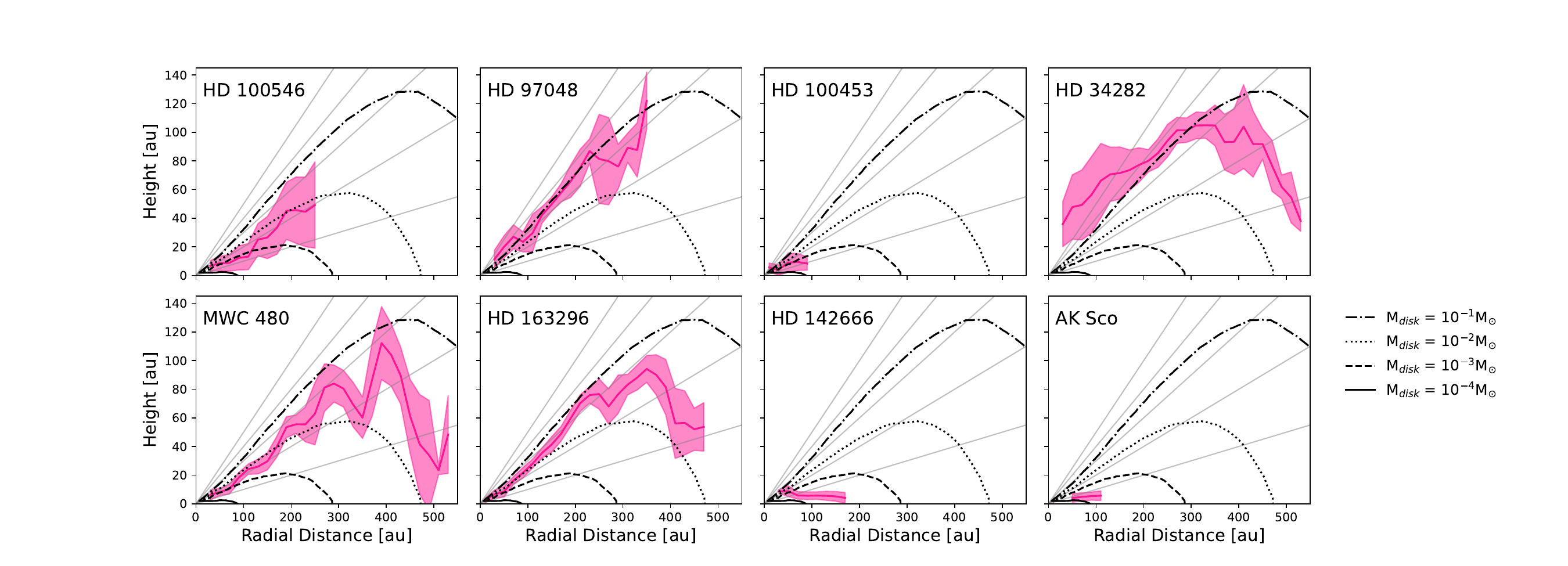}
      \caption{Same as Figure \ref{ttauri_obs}, but for the Herbig sample and models.
              }
         \label{herbig_obs}
\end{figure*}

Figures \ref{ttauri_obs} and \ref{herbig_obs} display the emission surfaces extracted from the CO channel map emission of each disk using ALFAHOR \citep{Paneque_2023_vert, Stapper_2023_Herbig_surf}. For comparison, the  location of the $\tau=1$ layer from the fiducial models using $R_c = 100$\,au of each sampled disk mass are plotted in each panel.

\section{CO isotopologue model fluxes using hydrostatic equilibrium solver}

When using the hydrostatic equilibrium solver in DALI, the disk density converges to a physical structure given by the equilibrium between the stellar gravitational potential and the thermal structure. The disk integrated fluxes may vary when comparing the hydrostatic equilibrium output to the models with static gaussian distributions. For optically thin emission in low-mass disks, the fluxes will be larger in the static models, which means the disk masses will be underestimated in comparison to those obtained from model fluxes when hydrostatic equilibrium is considered. In the case of high-mass disks the reverse happens, the static gaussian models underestimate the fluxes, therefore overestimate the disk masses.

\begin{figure*}[h!]
   \centering
   \includegraphics[width=\hsize]{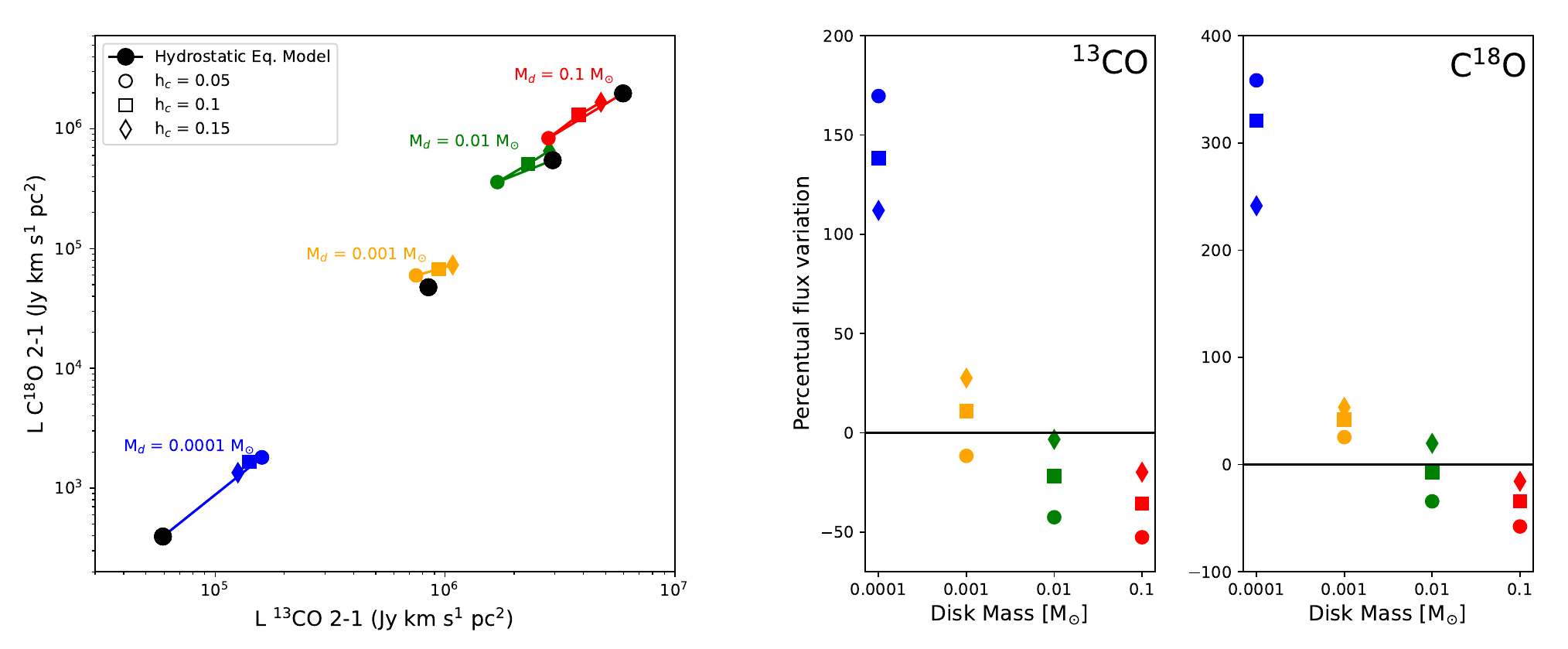}
      \caption{Comparison of retrieved fluxes from DALI models when using hydrostatic equilibrium and not. Left panel shows the $^{13}$CO and C$^{18}$O emission fluxes, black dots corresponds to models obtained after iterating and solving the hydrostatic equilibrium equations. Colored dots show the fluxes from models with a single iteration using a static gaussian material distribution. Right two panels display the percentual variation of the model flux from the static models compared to the one which uses hydrostatic equilibrium to determine its thermal and density structure.
              }
         \label{lucas_comp_fluxes}
\end{figure*}

Figure \ref{lucas_comp_fluxes} shows this effect displaying the total flux difference and percentual variation. This flux variation explains the generally higher masses of \citet{stapper_2024_gasmass} and the lower masses obtained by \citet{Miotello_2017} for MY Lup and GW Lup compared to our mass values assuming an ISM carbon abundance (see Fig. \ref{final_mass_plot}).

\end{appendix}

\end{document}